\documentclass[journal]{IEEEtran}
\usepackage{amsmath,amsfonts}
\usepackage{algorithmic}
\usepackage{algorithm}
\usepackage{array}
\usepackage[caption=false,font=normalsize,labelfont=sf,textfont=sf]{subfig}
\usepackage{textcomp}
\usepackage{stfloats}
\usepackage{url}
\usepackage{hyperref}
\usepackage{verbatim}
\usepackage{graphicx}
\usepackage{multirow}
\usepackage{cite}
\usepackage{color}
\usepackage{booktabs}
\definecolor{Revision}{RGB}{0, 0, 0}
\definecolor{Response}{RGB}{0, 0, 139}
\definecolor{Letter}{RGB}{0, 60, 255}
\definecolor{summary}{RGB}{0, 60, 255}

\begin{document}

\title{VisionUnite: A Vision-Language Foundation Model for Ophthalmology Enhanced with Clinical Knowledge}

\author{Zihan Li, Diping Song, Zefeng Yang, Deming Wang, Fei Li, Xiulan Zhang,\\ Paul E. Kinahan,~\IEEEmembership{Life Fellow,~IEEE,} Yu Qiao, ~\IEEEmembership{Senior Member,~IEEE}
\thanks{Zihan Li is with Shanghai Artificial Intelligence Laboratory, Shanghai, 200232, China and the University of Washington, Seattle, WA 98195, USA.}
\thanks{Diping Song is with Shanghai Artificial Intelligence Laboratory, Shanghai, 200232, China.}
\thanks{Yu Qiao is with Shanghai Artificial Intelligence Laboratory, Shanghai, 200232, China and Shenzhen Institutes of Advanced Technology, Chinese Academy of Sciences,  Shenzhen, 518055, China.}
\thanks{Zefeng Yang, Deming Wang, Fei Li, and Xiulan Zhang are with State Key Laboratory of Ophthalmology, Zhongshan Ophthalmic Center, Sun Yat-sen University, Guangdong Provincial Key Laboratory of Ophthalmology and Visual Science, Guangdong Provincial Clinical Research Center for Ocular Diseases, Guangzhou, 510060, China}
\thanks{Paul E. Kinahan is with the Department of Bioengineering and the Department of Radiology, University of Washington, Seattle, WA 98195, USA.}
\thanks{Zihan Li and Diping Song have equal contributions to this work.}
\thanks{Corresponding author: Yu Qiao}
}

\markboth{Journal of \LaTeX\ Class Files,~Vol.~14, No.~8, August~2021}%
{Shell \MakeLowercase{\textit{et al.}}: A Sample Article Using IEEEtran.cls for IEEE Journals}

\maketitle

\begin{abstract}
The need for improved diagnostic methods in ophthalmology is acute, especially in the underdeveloped regions with limited access to specialists and advanced equipment. Therefore, we introduce VisionUnite, a novel vision-language foundation model for ophthalmology enhanced with clinical knowledge. VisionUnite has been pretrained on an extensive dataset comprising 1.24 million image-text pairs, and further refined using our proposed MMFundus dataset, which includes 296,379 high-quality fundus image-text pairs and 889,137 simulated doctor-patient dialogue instances. Our experiments indicate that VisionUnite outperforms existing generative foundation models such as GPT-4V and Gemini Pro. It also demonstrates diagnostic capabilities comparable to junior ophthalmologists. VisionUnite performs well in various clinical scenarios including open-ended multi-disease diagnosis, clinical explanation, and patient interaction, making it a highly versatile tool for initial ophthalmic disease screening. VisionUnite can also serve as an educational aid for junior ophthalmologists, accelerating their acquisition of knowledge regarding both common and underrepresented ophthalmic conditions. VisionUnite represents a significant advancement in ophthalmology, with broad implications for diagnostics, medical education, and understanding of disease mechanisms. The source code is at \href{https://github.com/HUANGLIZI/VisionUnite}{https://github.com/HUANGLIZI/VisionUnite}.
\end{abstract}


\vspace{-4mm}
\section{Introduction}
\IEEEPARstart{I}{n} recent years, the global prevalence of ophthalmic diseases has surged beyond 2.2 billion, with over 1 billion individuals experiencing visual impairment due to limited access to essential medical services for conditions like myopia, hyperopia, glaucoma, and cataracts. The critical situation primarily stems from a shortage of ophthalmologists in low-income and middle-income regions, resulting in inadequate provision of ophthalmic services. Compounding this challenge, the World Health Organization's World Vision Report estimates a staggering cost of 14.3 billion to address these issues, underscoring the financial burden. Consequently, there is an escalating need for swift and precise comprehensive diagnoses facilitated by existing artificial intelligence (AI) technology. 

Notably, numerous researchers have made strides in developing vision models for diagnosing eye diseases, exemplified by works such as the works~\cite{de2018clinically, dai2021deep}. Current AI-based models for ophthalmology face three significant challenges: disease-specific diagnosis limitations, ineffective user interactions, and lack of result interpretability. Firstly, these models are often tailored to diagnose specific diseases and cannot provide comprehensive assessments for multiple conditions simultaneously. It is a critical shortfall, as patients frequently suffer from multiple ailments, particularly in older populations. For instance, as reported by the American Academy of Ophthalmology~\cite{Kern2015}, it is not uncommon for individuals aged 65 and above to have more than one eye disease. Secondly, there is a persistent issue with effective user interaction, which is essential for practical clinical implementation. Thirdly, many of these AI models lack interpretability in their diagnostic results, which is crucial for trust and reliability in medical settings.

An ideal solution would involve a comprehensive large vision-language model seamlessly managing diverse clinical scenarios, including disease screening, diagnostic process optimization, and junior ophthalmologist training. Such a model would integrate visual and linguistic data effectively, aligning closely with the diagnostic criteria used by medical professionals and adhering to clinical consensus guidelines. This approach would ideally involve identifying the lesion area and type before proceeding with a diagnosis, enhancing both the interpretability and accuracy of medical evaluation. However, prevailing models primarily fall short of this ideal. Current vision models like RETFound~\cite{zhou2023foundation} often diagnose diseases without explaining their findings, which lacks interpretability and deviates from medical standards. Advanced language models show a significant shortfall in effective vision-language integration~\cite{singhal2023large, li2023lvit}. This issue extends to poor user interaction and limits model responsiveness to user needs~\cite{moor2023foundation}. These limitations impair the accuracy and universality of diagnoses made by models, underscoring the need for an integrated approach that can combine visual and linguistic information.

\begin{figure*}[!ht]
\setlength{\abovecaptionskip}{-1.5mm}
  \centering
\includegraphics[width=\textwidth]{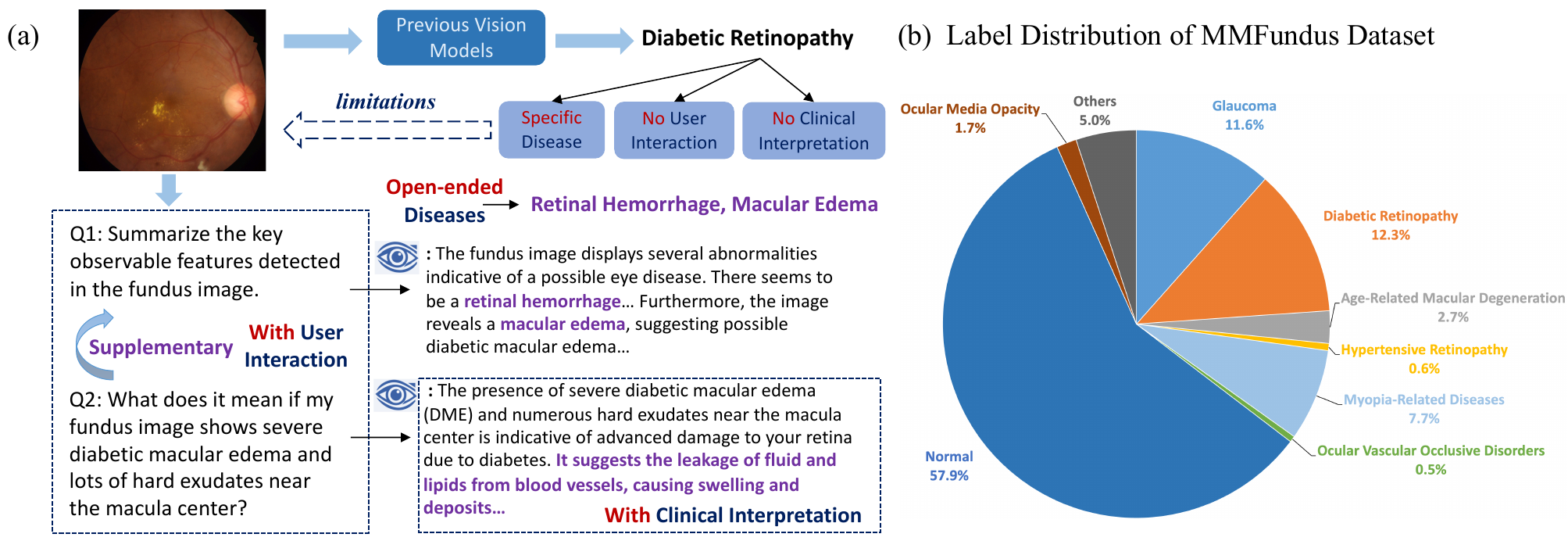}
\caption{\textbf{(a)} Previous vision models could only diagnose specific diseases as positive or negative, lacking the ability to provide clinical explanations or interact with patients. In contrast, our proposed VisionUnite represents a new paradigm. It can predict a wide range of diseases and allows real-time conversations with users, incorporating their feedback. Furthermore, VisionUnite offers clinically relevant explanations in its output, making it more understandable and useful. \textbf{(b)} The label distribution of the proposed MMFundus dataset, which includes eight main categories excluding the "Others" class.}
\label{Fig1}
\vspace{-6mm}
\end{figure*}

We introduce VisionUnite, a large vision language model tailored for ophthalmology, incorporating extensive clinical knowledge to effectively address these challenges. As depicted in Figure \ref{Fig1} (a), VisionUnite addresses three critical challenges: 1) the inability to predict open-ended multiple Diseases, 2) the lack of effective user interaction, and 3) diagnostic results with limited interpretability. Regarding the first challenge, VisionUnite demonstrates the capability to predict a wide range of lesion types in fundus images during the initial responses, encompassing conditions like Retinal Hemorrhage and Macular Edema. It stands in contrast to existing models constrained by data size and disease types, precluding open-ended disease prediction. Addressing the challenge of the absence of user interaction, VisionUnite implements multiple rounds of dialogue and effectively follows user instructions. The multiple rounds of dialogue enhance the comprehension of models to user commands and image information and contribute to the generation of more robust diagnostic reports. Finally, VisionUnite improves interpretability by providing detailed explanations of diagnoses, as illustrated in Figure \ref{Fig1}. These detailed responses help clinicians better understand the rationale behind the AI's decisions, thereby fostering greater trust in the model’s diagnostic outputs.

\begin{figure*}[!ht]
\setlength{\abovecaptionskip}{-1.5mm}
  \centering
\includegraphics[width=0.96\textwidth]{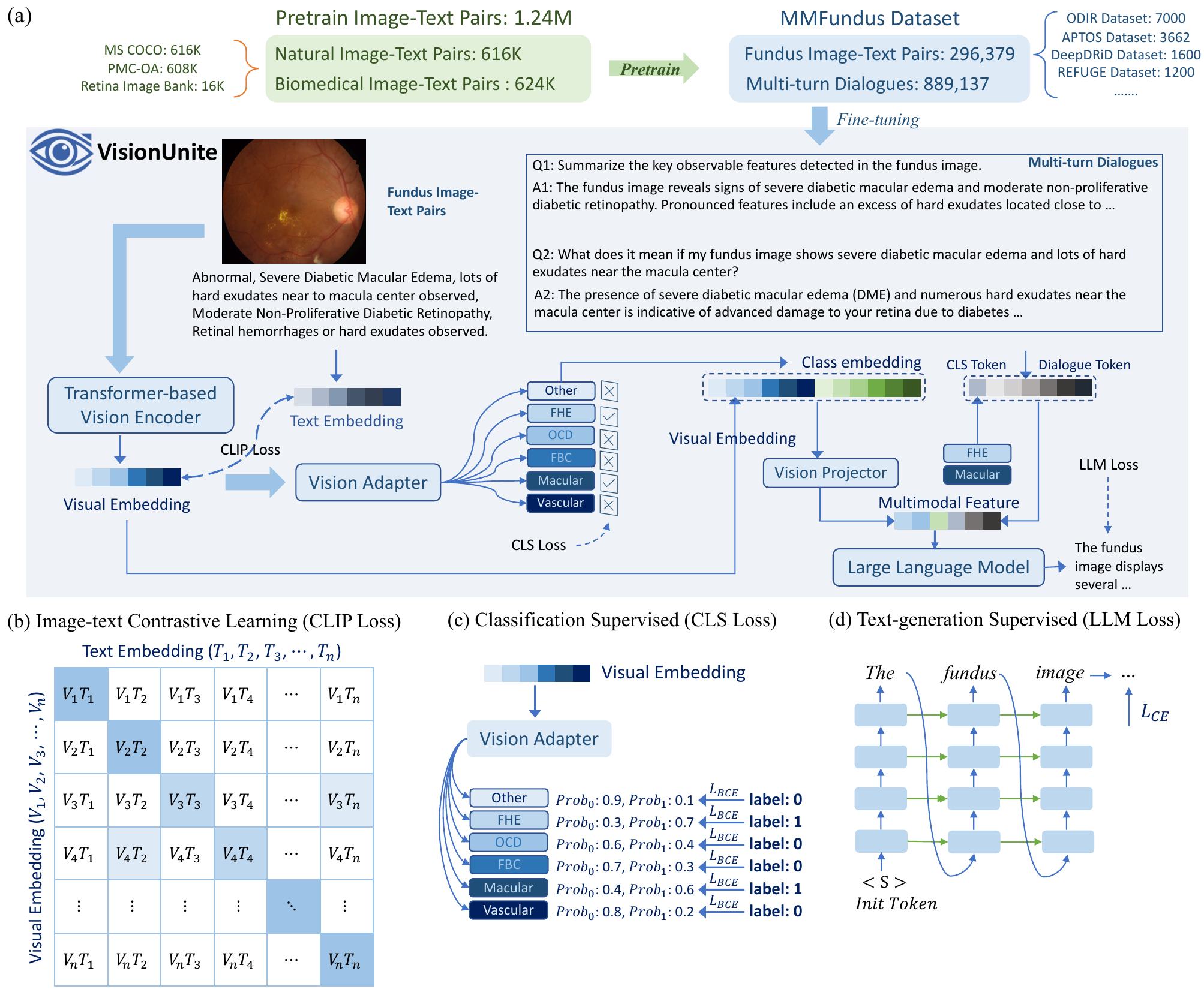}
\caption{\textbf{(a)} VisionUnite is built with a transformer-based vision encoder and a specialized vision adapter designed for classifying six different signs including Vascular, Macular, FBC (Fundus Boundary Color), OCD (Optical Cup Disc), FHE (Fundus Hemorrhages Examination), and Others. It includes a vision projector to align visual embeddings with text tokens. \textbf{(b)} The illustration of image-text contrastive learning (CLIP Loss). \textbf{(c)} The illustration of classification supervised learning (CLS Loss). \textbf{(d)} The illustration of text-generation supervised learning (LLM Loss).}
\label{Fig1-2}
\vspace{-6mm}
\end{figure*}

\vspace{-2mm}
\section{Related Work}
Vision-Language Foundation Models (VLFMs) have become a cornerstone in multimodal AI, skillfully blending vision and language domains to comprehend and generate multimodal content. These models, as discussed in references such as \cite{teamgemini, 2023GPT4VisionSC}, are trained on extensive datasets featuring images paired with textual annotations, facilitating high proficiency in tasks like image captioning, visual question answering, and cross-modal retrieval. The use of transformer architectures is central to their ability, allowing them to adeptly manage the intricacies of multimodal inputs \cite{zhang2024llama, touvron2023llama}. Recent innovations like the LLaMA-Adapter\cite{zhang2024llama} push these capabilities further by introducing adaptability in processing diverse datasets, enhancing the model's flexibility and application range. Similarly, LLaVa-Med\cite{li2024llava} extends these principles into the medical domain, integrating varied data sources to construct a comprehensive view of patient health, thus aiding better clinical decision-making.
Despite these advancements, significant challenges persist. The complexity inherent in VLFMs often results in a "black box" scenario, particularly problematic in healthcare settings where interpretability is crucial ~\cite{singhal2023large, moor2023foundation}. Models like Lama-adapter and Llava-med, while innovative, do not fully address the lack of transparency in how decisions are derived, which can be a barrier to their trust and utility in clinical practice. InternVL~\cite{chen2024internvl} represents a novel approach to generic visual-linguistic tasks. However, the complexity of the multi-stage training process, especially with layers of contrastive and generative training, can make it challenging to trace the specific decision process. Therefore, we propose the vision adapter to align coarse-grained labels explicitly. Additionally, there remains a significant gap in effective vision-language interaction. The current models still struggle to mimic the human interaction between visual cues and linguistic context, which limits the dynamic interactions.

\vspace{-2mm}
\section{Method}
\vspace{-2mm}
\subsection{Overview}
The architecture of VisionUnite is illustrated in Figure \ref{Fig1-2} (a), featuring four key components: a Transformer-based vision encoder, a Vision adapter, a vision
projector and a Large language model fine-tuned on LLaMA model~\cite{touvron2023llama}. For the pre-training of VisionUnite, we construct a comprehensive dataset of 1.19 million pretrained image-text pairs including natural image-text pairs~\cite{chen2015microsoft} and biomedical image-text pairs~\cite{PMCOA2003, lin2023pmc, RetinaBank}. The architecture of VisionUnite incorporates three training objectives: image-text contrastive learning, classification supervised learning, and text-generation supervised learning. These objectives help refine vision encoders, accurately categorize features, and guide the final text output. During pre-training, we use text-generation supervised learning as the primary objective to build robust connections between images and corresponding texts. Furthermore, we construct a Multimodal Fundus fine-tuning dataset (MMFundus) to train and evaluate the diagnostic performance of VisionUnite. The MMFundus dataset is currently the largest multimodal fundus dataset, including 296,379 sets of fundus image-text pairs and corresponding constructed 889,137 rounds of dialogue. The MMFundus dataset covers the eight main categories of fundus image as shown in Figure \ref{Fig1} (b) and its image part includes 53 datasets, such as ODIR~\cite{ODIR7000}, APTOS~\cite{APTOS3662}, DeepDRiD~\cite{DeepDRiD1600}, and REFUGE~\cite{REFUGE1200} dataset, etc. We design six sign categories including "Vascular", "Macular", "FBC (Fundus Boundary Color)", "OCD (Optical Cup Disc)", "FHE (Fundus Hemorrhages Examination)", and "Other" for each image. The six signs encompass all critical regions and functionalities of the eye, thereby providing a comprehensive assessment of eye health and disease.

\vspace{-6mm}
\subsection{Datasets Construction}
\vspace{-2mm}
\subsubsection{\textbf{Datasets for pre-training VisionUnite}}
In the pre-training phase of VisionUnite, we utilize a dataset comprising 1.24 million image-text pairs, encompassing around 616,435 natural image-text pairs sourced from the COCO dataset~\cite{chen2015microsoft} and around 623,816 biomedical image-text pairs sourced from the PMC-OA dataset~\cite{PMCOA2003, lin2023pmc} (608,027 images) and Retina Image Bank datasets~\cite{RetinaBank} (15,789 images). The expansive PMC-OA dataset encapsulates over 1.6 million image-text pairs drawn from 25 million scientific literature images and titles. Rigorous screening and preprocessing efforts are applied to distill this dataset down to 623,816 image-text pairs as outlined in the appendix. Throughout the pre-training process, we just employ text-generation supervised learning as the pre-training objective. This kind of strategy aims to establish robust representation connections between images and their corresponding texts. In the context of dialogue, the answer section is represented by captions from the text-image pairs. We have constructed 20 questions for the question section, which are categorized into two types: indicative long answers (10 sentences) and short answers (10 sentences). Depending on the length of the dialogue answer, a sentence from the relevant question type is selected as the interrogative component of the dialogue. The comprehensive list of the 20 constructed questions is provided in the appendix for reference.
\subsubsection{\textbf{MMFundus Dataset for fine-tuning VisionUnite}}
To realize the open-ended multi-disease diagnosis and comprehensive clinical explanations grounded in multiple rounds, we curate the pioneering multimodal fundus dataset, MMFundus. MMFundus dataset comprises 296,379 pairs of fundus images and corresponding text, accompanied by 889,137 rounds of dialogue. \textcolor{Revision}{While VisionUnite leverages substantial datasets for training, we develop a semi-automated annotation framework requiring minimal expert supervision. This framework employs a three-stage verification protocol: 1. Initial rule-based labeling leveraging structured report metadata. 2. Automated label propagation through classification by visual similarity. 3. Selective expert verification of boundary cases (5\% of the dataset). This methodology achieved 99\% annotation accuracy while reducing expert labeling needs by 95\% through an iterative self-training approach.} Drawing from 52 public datasets and one private dataset, the image data in MMFundus embodies a diverse spectrum of ophthalmic conditions. The textual and dialogue components are sourced through doctor annotations, inherent category labels, and the automated generation capabilities of large language models such as InternLM~\cite{team2023internlm} and GPT-4~\cite{achiam2023gpt}, as the details are provided in the appendix. For the testset, all image descriptions and dialogue data have undergone annotation or confirmation by medical professionals. In constructing the training set, we initially utilized doctor annotations and category labels to establish a prototype dataset. Leveraging the prototype dataset alongside a large language model, we automatically generate the image-related descriptions and dialogues batch by batch. We also design a hierarchical approach to generate descriptions at three levels—Normal/Abnormal, disease name, and clinical explanation based on the disease within the dataset. \textcolor{Revision}{Our method for ensuring the quality and accuracy of automatically generated content in the MMFundus dataset employs a comprehensive, multi-layered validation framework: (1) the expert-annotated prototype dataset establishing foundational parameters; (2) the hierarchical three-level annotation architecture with sign-level constraints that intrinsically validates semantic consistency; (3) an iterative refinement protocol where identified errors trigger complete batch regeneration until reaching 99\% accuracy.} As illustrated in Figure 2, the annotation "Abnormal" signifies that the image deviates from normalcy, while the descriptor "Severe Diabetic Macular Edema" denotes the specific disease depicted in the image. The statement "lots of hard exudates near the macula center observed, Moderate Non-Proliferative Diabetic Retinopathy, Retinal hemorrhages, or hard exudates observed" offers a detailed explanation of this disease. The inclusion of sign-level explanations is pivotal for accurate disease diagnosis. To innovate within this realm, we introduce corresponding sign-level labels to each image, namely Vascular, Macular, FBC (Fundus Boundary Color), OCD (Optic Cup Disc), FHE (Fundus Hemorrhages Exudation), and Other. Specifically, vascular pertains to ocular artery and vein-related conditions. Macular addresses diseases within or associated with the macular area. FBC involves signs affecting the boundary areas and the overall fundus background, such as the Leopard Fundus. OCD encompasses diseases related to the cup and disc. FHE focuses on diseases related to retinal hemorrhage and exudation. We generate the corresponding sign labels for each image using the original label information of fundus images based on  above standards. The distinction sets MMFundus apart from other multimodal datasets.
\vspace{-3mm}
\subsection{The architecture of VisionUnite}
We propose a large vision-language foundation model with clinical knowledge to enable the model to achieve open-ended prediction and have user interaction and clinical interpretation capabilities. Illustrated in Figure \ref{Fig1-2} (a), our proposed VisionUnite comprises a Transformer-based vision encoder, a vision adapter tailored for signs-level classification, a vision projector for aligning visual embeddings and text tokens, and a substantial language model fine-tuned on llama-7B. The large language model differs from the vision model in that it can achieve open-ended prediction. By leveraging the combination of image-text pairs and extensive dialogue information, VisionUnite excels in responding to user questions based on visual data. We introduce visual features with sign-level features into the model, which improve its ability to understand images and further enhance its clinical interpretation ability. This design can extend its ability to analyze fundus images for medical diagnosis. Our model is also designed for multi-round dialogues, which helps the model follow user instructions and achieve better problem-understanding ability.

\subsubsection{\textbf{Transformer-Based Vision Encoder}}
We leverage the EVA02 model~\cite{fang2023EVA} with the CLIP Vision Encoder to collectively serve as the Transformer-Based Vision Encoder. Our configuration incorporates 12 layers of EvaBlock within the original vision Transformer Block~\cite{dosovitskiy2020vit}. Notably, in our design, the GELU activation~\cite{hendrycks2016gaussian} in the original vision Transformer Block is replaced with the more advanced SwiGLU~\cite{dauphin2017language}. It is used to enhance the performance of the FFN (position feedforward network) layer in the Transformer architecture~\cite{zhang2024llama}. Specifically, images, text descriptions, and questions are fed into corresponding encoders to form visual embeddings, text embeddings, and dialogue tokens during the training phase. In this phase, we use contrastive learning to supervise the encoding formation of visual and text embeddings. 
The VisionUnite model uses a vision encoder to extract visual embeddings which are then used by a vision adapter to make classification predictions for specific signs. This process enhances diagnostic accuracy and efficiency by identifying and localizing signs like lesions, which helps narrow down diagnostic options and leads to more focused investigations. The model reduces the need for broad differential diagnoses and increases accuracy by concentrating on clinical signs.

\subsubsection{\textbf{Vision Adapter}}
The vision adapter in VisionUnite acts as a secondary encoder that takes the visual embeddings from the vision encoder and performs detailed classification into six sign category embeddings. These categories—Vascular, Macular, FBC, OCD, FHE, and Other—cover essential aspects of eye anatomy and pathology, directly correlating with common diagnostic criteria used in ophthalmology. Specifically, Vascular for blood flow, Macular for central vision, FBC for boundary conditions, OCD for optic nerve health (glaucoma), FHE for retinal bleeding and exudates, and Other for miscellaneous conditions. Then we concatenate the visual embeddings and sign category embeddings as inputs to the vision projector, which is shown in the following formulas. 
\vspace{-2mm}
\begin{eqnarray}
&V_{embed}^{'} = [V_{embed}, CLS_{1}(V_{embed}), ..., CLS_{6}(V_{embed})]\\
&Token_{CLS} = \sum_{i=1}^{6}Token(\mathbb{I}(CLS_{i}(V_{embed})>0)K_{i})
\end{eqnarray}
where $V_{embed}^{'}$ represents the combination of visual embedding $V_{embed}$ and each sign category embedding $CLS_{i}(V_{embed})$. $\mathbb{I}$ is the indicator function, which can help embed the corresponding keyword embedding ($K_{i}$) and concatenate them to form $Token_{CLS}$ based on the prediction of the signs $(CLS_{i}(V_{embed})$. It demonstrates that VisionUnite is designed to mimic the diagnostic approach of medical professionals by first identifying specific signs and then narrowing down to precise disease identification. It allows VisionUnite to align with human medical expert processes, enhancing diagnostics. \textcolor{Revision}{The sign-level classification provides an interpretable intermediate layer between raw visual features and final diagnoses too.}

\subsubsection{\textbf{Vision Projector}}
The vision projector is designed to integrate and synchronize the visual embeddings with dialogue tokens, facilitating a seamless multimodal interaction within the model. This component takes the concatenated visual and sign category embeddings and aligns them with the dialogue tokens by matching their feature dimensions. The inclusion of the $CLS$ token encapsulates the predicted sign-level features, which are crucial for generating relevant and context-aware text responses in the model, as shown in the formulas.
\vspace{-2mm}
\begin{eqnarray}
&V_{proj} = Prefix_q+Unsqueeze(V_{embed}^{'})\\
&F_{mm} = Attn(V_{proj}) + [Token_{CLS}, Token_{D}]
\end{eqnarray}
where $V_{proj}$ represents the projected feature of enhanced visual embedding $V_{embed}^{'}$. $Prefix_q$ represents the query prefix. The multimodal features $F_{mm}$ combine attention visual embeddings, $CLS$ tokens ($Token_{CLS}$), and dialogue tokens ($Token_{D}$). $F_{mm}$ are the inputs for training the final text generation response within the fine-tuned LLaMA model.
VisionUnite integrates multiple modalities to improve interaction and understanding in multi-round dialogues, enhancing its accuracy and relevance in responding to user queries. This integration boosts its diagnostic performance and utility in clinical settings, making it a valuable tool for medical image analysis crucial for accurate diagnosis and treatment planning.

\subsubsection{\textbf{Fine-tuned LLaMA Model}}
The Fine-tuned LLaMA Model is an enhanced language model based on the LLaMA-7B framework. The model has been specifically fine-tuned to integrate seamlessly with vision components, enabling advanced text generation, classification, and open-ended predictive capabilities. The aligned visual-text data is fed into the fine-tuned LLaMA model, which uses the input to generate textual output that is contextually relevant to the visual input. Unlike typical vision models, the fine-tuned LLaMA model supports interactive dialogues and dynamic user interactions. \textcolor{Revision}{The multi-round dialogue capability allows clinicians to interrogate the reasoning process.}

\vspace{-4mm}
\subsection{The training objectives of VisionUnite}
In the architecture of VisionUnite, we have crafted three distinct training objectives to enhance convergence: image-text contrastive learning, classification supervised learning, and text-generation supervised learning. The utilization of image-text contrastive learning facilitates the refinement of visual encoders, aiding them in more effectively aligning fundus image features. Meanwhile, the application of classification-supervised learning contributes by furnishing accurate feature categories for both visual embeddings and dialogue tokens. We utilize the accuracy to guide the model training process, enhancing its overall performance. Finally, text-generation supervised learning plays a role in guiding the output of the language model, which is pivotal for achieving accurate and open-ended disease diagnoses. Unlike previous vision models, large language models are trained on vast amounts of diverse textual data. The extensive textual data enables them to generate context-aware responses and detailed explanations that are coherent and clinically relevant. We can achieve open-ended disease diagnosis using the generated text from language models. In contrast, the prediction categories of vision models are limited and concentrated on specific diseases.

\subsubsection{\textbf{Image-text contrastive learning}} To attain seamless alignment between image and text features, we employ image-text contrastive learning. We utilize LLaMA~\cite{touvron2023llama} SentencePiece tokenizer to get the text embedding, which ensures consistency with other training objectives. We leverage the CLIP loss to quantify the similarity between image embedding and text embedding, as illustrated in Figure \ref{Fig1-2} (b) and below:
\vspace{-1mm}
\begin{eqnarray}
&L_{CLIP}=(L_{img} + L_{text})/2\\
&L_{img} = -\frac{1}{N}{\sum_{i=1}^{N}[t_{i}\cdot log(p_{img,i})]}\\
&L_{text} =-\frac{1}{N}{\sum_{i=1}^{N}[t_{i}\cdot log(p_{text,i})]}
\end{eqnarray}
where $N$ denotes the number of samples in each batch. $p_{img, i}$ is the cosine similarities of image $i$ to all $N$ text embeddings and  $p_{text, i}$ is the cosine similarities of text $i$ to all $N$ image embeddings. $t_{i}$ denotes the soft label representation $\{p_{1},p_{2},p_{3},...,p_{i},...p_{n}\}$ of the corresponding image and text pairs along which the cross-entropy loss is computed.

\subsubsection{\textbf{Classification supervised learning}} In facilitating the acquisition of sign-level features, we employ conventional classification learning to guide the training of VisionUnite. Notably, we embrace multi-label classification while acknowledging that each sample may encompass more than one predicted category. This approach aligns more closely with the intricacies of real-world clinical scenarios, where patients may concurrently experience multiple types of eye diseases. For each category under supervision, we apply cross-entropy loss, aggregating these individual losses to derive the ultimate classification loss, as depicted in Figure \ref{Fig1-2} (c) and below:
\vspace{-1mm}
\begin{eqnarray}
L_{CLS}&=\sum_{
k=1}^{M}{L_{CLS,k}}
\end{eqnarray}
\vspace{-7mm}
\begin{eqnarray}
\begin{aligned}
{L_{CLS,k}}&=-\frac{1}{N}{\sum_{i=1}^{N}y_{k,i}\cdot log(p_{k,i})}\\
&-\frac{1}{N}{\sum_{i=1}^{N}}{(1-y_{k,i})\cdot log(1-p_{k,i})}
\end{aligned}
\end{eqnarray}
where $M$ and $N$ denote the number of categories and samples respectively. $L_{CLS, k}$ signifies the cross-entropy loss for category $K$. And $y_{k, i}$ denotes whether sample $i$ belongs to class $K$, and $p_{k, i}$ represents the probability of sample $i$ in class $K$.

\subsubsection{\textbf{Text-generation supervised learning}} Within the framework of VisionUnite, we employ text-generation supervised learning to guide the text output of LLM, which is an essential aspect given that the generated text must intuitively articulate the diagnostic results and their underlying rationale. The LLM loss aims to train the model to generate text closely resembling patterns in its training data as shown in Figure \ref{Fig1-2} (d), and its formulation is articulated as follows:
\vspace{-1mm}
\begin{eqnarray}
L_{LLM} = -\frac{1}{N}\sum_{i=1}^{N}\sum_{j=1}^{T_i}\log p_{\theta}(t_{i,j}|t_{i,<j},x_i)
\end{eqnarray}
where $N$ represents the number of samples in the training set, $T_i$ denotes the target sequence length for the i-th sample, and $t_{i,j}$ signifies the j-th target word in the i-th sample. $t_{i,<j}$ corresponds to the preceding j-1 target words in the i-th sample. $x_i$ is the input sequence for the i-th sample, and $p_{\theta}(t_{i,j}|t_{i,<j},x_i)$ denotes the probability assigned by the model to the j-th target word in the i-th sample, with $\theta$ representing the model parameters.

\vspace{-2mm}
\section{Experiments}
\subsection{Experimental Settings}
\subsubsection{\textbf{The image source of MMFundus dataset}}
The large-scale multimodal fundus dataset (MMFundus) is constructed with images from 53 datasets. The private dataset is from a hospital in China mainland, which includes two major diseases ``myopia" and ``glaucoma". The distribution is in Table. \ref{datadis}.

\vspace{-5mm}
\begin{table}[!ht]
\Large
\setlength{\abovecaptionskip}{0mm}
  \centering
  \caption{The data distribution of MMFundus dataset}
  \resizebox{\columnwidth}{!}{%
    \begin{tabular}{c|c|c|c|c|c}
    \toprule
    \textbf{Dataset} & Num & \textbf{Dataset} & Num & \textbf{Dataset} & Num \\
    \midrule
    HRF~\cite{HRF45} & 45 & INSPIRE-AVR~\cite{INSPIRE-AVR40} & 40 & IOSTAR~\cite{IOSTAR30}& 30\\
   RITE~\cite{RITE40}& 40 & G1020~\cite{G1020} & 1020 & GAMMA~\cite{GAMMA100}& 100 \\ 
    ORIGA~\cite{ORIGA650}& 650& REFUGE~\cite{REFUGE1200}& 1200 & ODIR~\cite{ODIR7000}&7000 \\
    PALM~\cite{PALM1200}& 1200& RFMiD~\cite{RFMiD3200}& 3200 & RFMiDv2~\cite{RFMiD2_860}& 860\\
    APTOS~\cite{APTOS3662}& 3662 &DeepDRiD~\cite{DeepDRiD1600}& 1600 & EyePACS~\cite{EyePACS35126} & 35126\\
    IDRID~\cite{IDRID516}&516& ADAM~\cite{ADAM1200}&1200&ACRIMA~\cite{ACRIMA705}&705\\
    MESSIDOR-2~\cite{MESSIDOR2_1748}&1748&JSIEC~\cite{JSIEC_1000}&1000& DeepEyeNet~\cite{EYENET_15708}&15708\\
    LAG~\cite{LAG_4854}&4854&PARAGUAY~\cite{PARAGUAY_757}&757&PAPILA~\cite{PAPILA_488}&488\\
    STARE~\cite{STARE_397}&397&FIVES~\cite{FIVES_800}&800&FUND~\cite{FUND_179}&179\\
    E-ophta~\cite{E-ophta_463}&463&BRSET~\cite{BRSET_16266}&16266&MuReD~\cite{MuReD_2208}&2208\\
    OIA-DDR~\cite{OIA-DDR_12522}&12522&SUSTech-SYSU~\cite{SUSTech-SYSU_1219}&1219&Cataract~\cite{Cataract_601}&601\\
    DGOCF~\cite{DGOCF_9939}&9939&BoVW~\cite{BoVW_2013}&2013 &HarvardGlaucoma~\cite{HarvardGlaucoma_1544}&1544\\
    RIM-ONE~\cite{RIM-ONE_485}&485&CHAKSU~\cite{CHAKSU_1345}&1345&DiaRetDB~\cite{DiaRetDB_89}&89\\
    LSD~\cite{LSD_175}&175&GNG~\cite{GNG_400}&400&AOD~\cite{AOD_14813}&14813\\
    DHRF~\cite{DHRF_2757}&2757&VietAI~\cite{VietAI_3435}&3435&ToxoFundus~\cite{ToxoFundus_411}& 411\\
    Papilledema~\cite{Papilledema_1369}&1369& BEH~\cite{BEH_634}&634&WilliamHoyt~\cite{WilliamHoyt_856}&856\\
    ROI~\cite{ROI_1120}&1120&ROD~\cite{ROD_281}&281&BiDR~\cite{BiDR_2838}&2838\\
     AIROGS~\cite{AIROGS_101442}&101442&Private&33029 \\
    \midrule
    Summary & \multicolumn{5}{|c}{296,379 images}\\
    \bottomrule
    \end{tabular} 
    }
  \label{datadis}%
\end{table}%
\vspace{-3mm}

\subsubsection{\textbf{Baseline methods}}
We employ three distinct baseline methods for comparative analysis with VisionUnite, GPT-4V~\cite{2023GPT4VisionSC}, Gemini Pro~\cite{teamgemini}, and manual interpretation from the junior ophthalmologist. All methods are evaluated on a test set comprising 180 images. The manual interpretation involves diagnostic opinions and bases generated by junior ophthalmologists. The outcomes from three models and junior ophthalmologists undergo an assessment by senior ophthalmologists to evaluate the accuracy and relevance of the generated text.

\subsubsection{\textbf{Evaluation and statistical analysis}}
We evaluate model efficacy using a multi-round VQA dataset from the MMFundus dataset, consisting of 180 samples and 540 question rounds. This dataset includes a spectrum of fundus conditions, from healthy to thirteen different ophthalmic diseases. Our evaluation benchmarks model performance against open-set clinical diagnostics based on two key criteria: diagnostic accuracy and relevance. Diagnostic accuracy measures the precision of disease identification and categorization. However, since accuracy alone may not fully capture a model's performance, we introduce diagnostic relevance to assess the quality of model responses and distinguish between different models. Our testing setup involves presenting a fundus image and three related questions to a junior ophthalmologist with one year of experience and eliciting responses from three advanced vision-language models: GPT-4V~\cite{2023GPT4VisionSC} (gpt-4-1106-vision-preview), Gemini Pro~\cite{teamgemini} (gemini-pro-vision), and our proposed VisionUnite. These responses are then assessed by three senior ophthalmologists with over ten years of experience, focusing on both diagnostic accuracy and relevance. We statistically calculate the average diagnostic relevance to compare the different models. Additionally, we analyze model misdiagnoses, categorizing errors as minor or major based on their severity. We also compute the 95\% confidence intervals and p-values for each model's results, particularly comparing them to our VisionUnite model, ensuring that responses are integrated into coherent paragraphs without additional AI-generated content for fair evaluation.
We also design a multiple-choice evaluation experiment to evaluate the diagnostic performance across various models in a close-set setting. We conduct experiments on 2233 cases and calculate the multiple-choice accuracy of ten diseases.  
The details are as follows:

\textbf{1. Diagnostic accuracy:} We follow the clinical evaluation standards. 1. In cases of extra answers, we consider them to be incorrect answers. 2. In case of just missing unnecessary diagnostic information, we still consider the responses to be correct. 3. In our evaluation, the answer must only include all diagnosable diseases to be considered correct. We use the Wilson method~\cite{wilson1927probable} to estimate 95\% confidence interval of diagnostic accuracy and calculate the p-value with the two-sided T-test and the above Wilson estimation.

\textbf{2. Diagnostic relevance:} Senior ophthalmologists rank the response sets based on their alignment with the accurate diagnosis, scoring the most consistent response as four and the least as one. The diagnostic relevance is designed to refine performance evaluation by taking into account factors that mere diagnostic accuracy might miss. For example, a response could be accurate in diagnostic prediction yet fail to adhere strictly to diagnostic criteria, which will diminish its overall diagnostic relevance. The bootstrap~\cite{diciccio1996bootstrap} method is used to estimate the 95\% confidence interval of diagnostic relevance and the two-sided T-test is applied to calculate its p-value.

\textcolor{Revision}{\textbf{3. Sentence-BERT (SBERT) Similarity}: Our evaluation also employs Sentence-BERT (SBERT) Similarity \cite{reimers-2019-sentence-bert} as the quantitative metric for assessing semantic correspondence between generated responses and ground truth. We report the results of SBERT Similarity across 540 multi-round dialogue instances.}

\textbf{4. Multiple-choice accuracy:} We provide four distinct options from all disease labels in the MMFundus dataset that models should select for the most likely diagnostic outcome based on the provided fundus image. The other three options are randomly selected except the correct options.

\vspace{-3mm}
\subsection{Comparison of Diagnosis between Ophthalmologist and Large Vision-Language Models}
In our study, we conduct a comprehensive assessment of the diagnostic capabilities exhibited by the junior ophthalmologist compared with models, which include Gemini Pro, GPT-4V, and our proposed VisionUnite. The evaluation is performed on the designated test dataset. Additionally, we explore an examination of the diagnostic relevance between the responses provided by the junior ophthalmologist and those generated by models. As shown in Table \ref{Fig2}, the overall diagnostic accuracy of VisionUnite is over 45\% and 28\% higher than Gemini Pro and GPT-4V respectively, and the corresponding p-value is less than 0.001. The diagnostic relevance of VisionUnite is also over 62\% and 25\% higher than Gemini Pro and GPT-4V respectively with the p-value less than 0.001. Compared to the junior ophthalmologist, the overall accuracy of VisionUnite is approximately 4.5\% higher with the p-value being 0.0876, and the diagnostic relevance is 0.75\% higher (from 2.915 to 2.937). In each round of result evaluation, we find that the performance of VisionUnite in the first round of diagnostic Q\&A is higher than that of junior ophthalmologist, with an accuracy rate of approximately 7.2\% and a diagnostic relevance of 2.6\%. The result indicates that VisionUnite has stronger analytical and reasoning abilities for fundus images, thus achieving superior performance in the first round of diagnosis. In the second and third rounds of diagnostic Q\&A, the performance of VisionUnite is comparable to that of a junior ophthalmologist and far superior to GPT-4V and Gemini Pro. It also indicates that VisionUnite has the same problem-solving and interpretation abilities as doctors.

\begin{table*}[t]
\centering
\LARGE
\caption{Comparison of Disease Diagnosis Accuracy and Relevance between Ophthalmologist and Models. All p-values are calculated with the two-sided T-test between VisionUnite and others. * means p-value $<$ 0.1 and ** means p-value $<$ 0.001.}
\vspace{-2.5mm}
\resizebox{\textwidth}{!}{
\begin{tabular}{l|c|c|c|c|c|c|c|r}
\toprule
\multirow{2}{*}{Method}& \multicolumn{2}{c|}{Gemini Pro~\cite{teamgemini}} & \multicolumn{2}{c|}{GPT-4V~\cite{2023GPT4VisionSC}} & \multicolumn{2}{c|}{VisionUnite} & \multicolumn{2}{c}{Doctor} \\
& \textit{Accuracy (95\% CI)} &\textit{Relevance (95\% CI)}& \textit{Accuracy (95\% CI)} &\textit{Relevance (95\% CI)} & \textit{Accuracy (95\% CI)} &\textit{Relevance (95\% CI)}& \textit{Accuracy (95\% CI)} &\textit{Relevance (95\% CI)}\\
\midrule
Round 1&.194 (.143 \textit{to} .258) **&1.572 (1.467 \textit{to} 1.700) **&.406	(.337 \textit{to} .479) **&2.200 (2.072 \textit{to} 2.333) **&\textbf{.750 (.682 \textit{to} .808)}&\textbf{3.172 (3.050 \textit{to} 3.289)}&.678 (.606 \textit{to} .742)&3.056 (2.906 \textit{to} 3.206)\\
Round 2&.278 (.218 \textit{to} .347) **& 1.828 (1.683 \textit{to} 1.978) ** & .444 (.374 \textit{to} .517) **&2.300	(2.161 \textit{to} 2.433) **&\textbf{.794 (.730 \textit{to} .847)}& \textbf{2.961 (2.822 \textit{to} 3.106)} & .756	(.688 \textit{to} .813)&2.911	(2.750 \textit{to} 3.067)\\
Round 3& .506 (.433	\textit{to} .578) **& 2.039	(1.889 \textit{to} 2.206) **& .644 (.572 \textit{to} .711) * & 
2.506 (2.372 \textit{to} 2.639) &\textbf{.794 (.730 \textit{to} .847)} &2.678 (2.511 \textit{to} 2.839)& .772 (.706 \textit{to} .827) &\textbf{2.778 (2.594 \textit{to} 2.956)}\\
\midrule
Mean $\uparrow$&.326 (.288 \textit{to} .367) **&1.813 (1.726 \textit{to} 1.894) **&.498 (.456 \textit{to} .540) **&2.335 (2.256 \textit{to} 2.413) **&\textbf{.780	(.743 \textit{to} .813)}&\textbf{2.937 (2.854 \textit{to} 3.020)}&.735 (.696 \textit{to} 0.771) * &2.915	(2.826 \textit{to} 3.004)\\
\bottomrule
\end{tabular}
}
\label{Fig2}
\vspace{-4mm}
\end{table*}

\begin{table*}[!ht]
\centering
\LARGE
\caption{\textcolor{Revision}{Comparison of close source API-based and open source fine-tuned methods on multi-rounds VQA using SBERT similarity (\%).}}
\vspace{-2.5mm}
\resizebox{\linewidth}{!}{
\begin{tabular}{c|cc|cccccccc|c}
\toprule
\multirow{2}{*}{Method}& \multicolumn{2}{c|}{\textbf{Close source API-based}}  & \multicolumn{8}{c|}{\textbf{Open source fine-tuned}} & \multirow{2}{*}{Doctor}\\
& Gemini Pro~\cite{teamgemini} & GPT-4V~\cite{2023GPT4VisionSC} & InstructBLIP~\cite{instructblip} & Mini-Gemini~\cite{li2024mini} & Qwen-VL~\cite{bai2023qwen} & InternVL~\cite{chen2024internvl} & LLaVA~\cite{liu2024visual} & Med-Flamingo~\cite{moor2023med} & LLaVA-Med~\cite{li2024llava} & VisionUnite & \\
\midrule
 Round 1 & 65.27 & 64.57 & 62.39 & 65.93 & 70.43 & 75.71 & 76.26 & 64.31 & 67.83 & \textbf{83.46} & 76.52 \\
 Round 2 & 63.90 & 69.56 & 61.78 & 65.52 & 71.18 & 73.02 & 73.54 & 63.79 & 68.11 & \textbf{78.53} & 72.11 \\
 Round 3 & 62.98 & 69.80 & 60.03 & 64.85 & 70.25 & 71.86 & 72.98 & 62.14 & 67.06 & \textbf{77.82} & 71.47 \\ \midrule
 Overall $\uparrow$ & 64.05 & 67.98 & 61.40 & 65.43 & 70.62 & 73.53 & 74.26 & 63.41 & 67.67 & \textbf{79.94} & 73.37\\
\bottomrule
\end{tabular}
}
\label{ExtendedR2}
\vspace{-4mm}
\end{table*}

\begin{table*}[!ht]
\centering
\LARGE
\caption{\textcolor{Revision}{Comparison of close source API-based methods and open source fine-tuned methods on multiple-choice VQA using accuracy (\%). ``Overall" represents overall performance of benchmark.}}
\vspace{-2.5mm}
\resizebox{\linewidth}{!}{
\begin{tabular}{c|cc|cccccccc}
\toprule
\multirow{2}{*}{Method}& \multicolumn{2}{c|}{\textbf{Close source API-based}}  & \multicolumn{8}{c}{\textbf{Open source fine-tuned}} \\
& Gemini Pro~\cite{teamgemini} & GPT-4V~\cite{2023GPT4VisionSC} & InstructBLIP~\cite{instructblip} & Mini-Gemini~\cite{li2024mini} & Qwen-VL~\cite{bai2023qwen} & InternVL~\cite{chen2024internvl} & LLaVA~\cite{liu2024visual} & Med-Flamingo~\cite{moor2023med} & LLaVA-Med~\cite{li2024llava} & VisionUnite \\
\midrule
 AMD & 77.78 & 73.10 & 80.17 & 76.61 & 81.87 & 81.29 & 83.04 & 25.73 & 16.37 & 85.38 \\
 AR & 0.00 & 0.00 & 0.00 & 0.00 & 0.00 & 0.00 & 0.00 & 100.00 & 100.00 & 0.00 \\
 BRVO & 43.75 & 50.00 & 50.00 & 43.75 & 43.75 & 37.50 & 50.00 & 31.25 & 31.25 & 50.00  \\ 
 Cataract & 70.00 & 80.00 & 80.00 & 85.00 & 75.00 & 85.00 & 90.00 & 30.00 & 15.00 & 95.00\\ 
 Chorioretinitis & 0.00 & 0.00 & 0.00 & 0.00 & 0.00 & 0.00 & 0.00 & 0.00 & 0.00 & 0.00\\
 CN & 25.00 & 25.00 & 75.00 & 25.00 & 25.00 & 25.00 & 25.00 & 25.00 & 0.00 & 25.00\\
 CRVO & 9.09 & 18.18 & 45.45 & 27.27 & 9.09 & 36.36 & 9.09 & 63.64 & 45.45 & 36.36\\
 CSR & 42.86 & 57.14 & 0.00 & 14.29 & 28.57 & 71.43 & 42.86 & 57.14 & 42.86 & 85.71\\ 
 DR & 79.19 & 84.56 & 76.51 & 79.87 & 80.54 & 94.63 & 87.25 & 36.91 & 26.85 & 98.66\\
 Drusen & 33.33 & 33.33 & 36.67 & 23.33 & 26.67 & 43.33 & 36.67 & 20.00 & 16.67 & 46.67\\
 Glaucoma & 74.07 & 78.40 & 59.30 & 67.90 & 78.40 & 89.51 & 91.36 & 24.07 & 25.31 & 100.00\\
 Health & 67.53 & 78.09 & 51.57 & 61.46 & 79.89 & 85.73 & 88.65 & 74.16 & 91.46 & 97.87\\
 HR & 0.00 & 0.00 & 0.00 & 0.00 & 0.00 & 33.33 & 0.00 & 33.33 & 33.33 & 0.00\\
 DME & 74.14 & 79.31 & 63.79 & 60.34 & 84.48 & 87.93 & 93.10 & 24.14 & 16.67 & 100.00\\ 
 MH & 45.16 & 48.39 & 16.13 & 38.71 & 54.84 & 64.52 & 48.39 & 22.58 & 25.81 & 67.74\\
 Myopia & 71.24 & 73.89 & 44.25 & 58.41 & 76.55 & 88.05 & 88.50 & 18.58 & 23.89 & 96.46\\
 No AMD & 73.63 & 77.17 & 54.34 & 63.02 & 82.96 & 85.85 & 90.68 & 52.73 & 66.56 & 100.00\\
 No DME & 81.82 & 90.91 & 63.64 & 72.73 & 72.73 & 81.82 & 90.91 & 36.36 & 36.36 & 100.00\\
 No Glaucoma & 50.00 & 91.67 & 41.67 & 66.67 & 75.00 & 91.67 & 100.00 & 16.67 & 16.67 & 100.00\\
 ODC & 68.75 & 68.75 & 65.63 & 62.50 & 56.25 & 87.50 & 62.50 & 43.75 & 21.88 & 87.50\\
 ODE & 27.27 & 54.55 & 27.27 & 36.36 & 27.27 & 45.45 & 18.18 & 36.36 & 27.27 & 45.45\\
 ODP & 50.00 & 50.00 & 50.00 & 50.00 & 50.00 & 50.00 & 50.00 & 0.00 & 0.00 & 50.00\\
 Other & 76.00 & 86.00 & 58.00 & 66.00 & 84.00 & 90.00 & 90.00 & 28.00 & 28.00 & 100.00\\
 Retinitis & 44.44 & 22.22 & 11.11 & 33.33 & 22.22 & 44.44 & 44.44 & 22.22 & 33.33 & 55.56\\
 Tessellation & 58.33 & 33.33 & 41.67 & 33.33 & 25.00 & 66.67 & 58.33 & 8.33 & 16.67 & 75.00\\
\midrule 
 \textbf{Overall $\uparrow$}& 69.19 & 75.28 & 55.17 & 62.65 & 76.80 & 84.33 & 85.22 & 49.17 & 56.47 & \textbf{94.36}\\
\bottomrule
\end{tabular}
}
\label{ExtendedFig5}
\vspace{-3mm}
\end{table*}

\begin{table*}[!ht]
\centering
\LARGE
\caption{Diagnosis comparison of specific Disease. * means p-value $<$ 0.1 and ** means p-value $<$ 0.001.}
\vspace{-3mm}
\resizebox{\textwidth}{!}{
\begin{tabular}{l|c|c|c|c|c|c|c|r}
\toprule
\multirow{2}{*}{Method}& \multicolumn{2}{c|}{Gemini Pro~\cite{teamgemini}} & \multicolumn{2}{c|}{GPT-4V~\cite{2023GPT4VisionSC}} & \multicolumn{2}{c|}{VisionUnite} & \multicolumn{2}{c}{Doctor} \\
& \textit{Accuracy (95\% CI)} &\textit{Relevance (95\% CI)}& \textit{Accuracy (95\% CI)} &\textit{Relevance (95\% CI)} & \textit{Accuracy (95\% CI)} &\textit{Relevance (95\% CI)}& \textit{Accuracy (95\% CI)} &\textit{Relevance (95\% CI)}\\
\midrule
AMD &.350 (.271 \textit{to} .439) **& 1.808 (1.633 \textit{to} 1.975) ** & .567 (.477 \textit{to} .652) ** & 2.375 (2.208 \textit{to} 2.542) ** & \textbf{.817 (.738 \textit{to} 0.876)} & \textbf{2.975 (2.792 \textit{to} 3.158)} & .625 (.536 \textit{to} .706) ** & 2.842 (2.642 \textit{to} 3.033)\\
BRVO & .143	(.050 \textit{to} .346) * & 1.238 (1.095 \textit{to} 1.429) ** & .429 (.245 \textit{to} .635) & 2.286 (1.952 \textit{to} 2.619) * & .571 (.365 \textit{to} .755) & 2.762 (2.381 \textit{to} 3.143) & \textbf{.952 (.773 \textit{to} .992)} & \textbf{3.714 (3.524 \textit{to} 3.905)} \\
Cataract& .333 (.097 \textit{to} 0.700) & 2.000 (1.333 \textit{to} 2.667) & .500 (.188 \textit{to} .812) & 2.667 (1.833 \textit{to} 3.500) & \textbf{.500 (.188 \textit{to} .812)} &2.500 (1.500 \textit{to} 3.500) & .500 (.188 \textit{to} .812) & \textbf{2.833 (2.000 \textit{to} 3.667)}\\
DR & .238 (.160 \textit{to} .339) ** & 1.702 (1.524 \textit{to} 1.905) ** & .393 (.295 \textit{to} .500) ** & 2.286 (2.095 \textit{to} 2.488) ** & \textbf{.679 (.573 \textit{to} .769)} & 2.964 (2.774 \textit{to} 3.190) & .536 (.430 \textit{to} .638) * & \textbf{3.048 (2.810 \textit{to} 3.274)}\\ 
Drusen & .167 (.030 \textit{to} .564) * & 1.667 (1.167 \textit{to} 2.333) * & .500 (.188 \textit{to} .812) & 2.333 (1.667 \textit{to} 2.833) * & \textbf{.833 (.436 \textit{to} .970)} & \textbf{3.500 (2.833 \textit{to} 4.000)} & .500 (.188 \textit{to} .812) & 2.500 (1.500 \textit{to} 3.500) \\
Glaucoma & .188	(.114 \textit{to} .296) ** & 1.870 (1.652 \textit{to} 2.130) ** & .188 (.114 \textit{to} .296) ** & 2.145 (1.957 \textit{to} 2.319) ** & \textbf{.739 (.625 \textit{to} .828)} & 2.971 (2.710 \textit{to} 3.203) & .725 (.610 \textit{to} 0.816) & \textbf{3.014 (2.739 \textit{to} 3.275)}\\
Myopia & .030 (.005 \textit{to} .153) ** & 1.424 (1.212 \textit{to} 1.667) **& .061 (.017 \textit{to} .196) ** & 1.848 (1.667 \textit{to} 2.030) ** & .697 (.527 \textit{to} .826) & 2.939 (2.636 \textit{to} 3.212) & \textbf{1.000 (0.896 \textit{to} 1.000)} & \textbf{3.788 (3.636 \textit{to} 3.909)}\\
\bottomrule
\end{tabular}
}
\label{Fig3}
\vspace{-3.5mm}
\end{table*}

\begin{table*}[!t]
\centering
\LARGE
\caption{Diagnosis comparison of specific sign. * means p-value $<$ 0.1 and ** means p-value $<$ 0.001.}
\vspace{-3mm}
\resizebox{\textwidth}{!}{
\begin{tabular}{l|c|c|c|c|c|c|c|r}
\toprule
\multirow{2}{*}{Method}& \multicolumn{2}{c|}{Gemini Pro~\cite{teamgemini}} & \multicolumn{2}{c|}{GPT-4V~\cite{2023GPT4VisionSC}} & \multicolumn{2}{c|}{VisionUnite} & \multicolumn{2}{c}{Doctor} \\
& \textit{Accuracy (95\% CI)} &\textit{Relevance (95\% CI)}& \textit{Accuracy (95\% CI)} &\textit{Relevance (95\% CI)} & \textit{Accuracy (95\% CI)} &\textit{Relevance (95\% CI)}& \textit{Accuracy (95\% CI)} &\textit{Relevance (95\% CI)}\\
\midrule
FHE & .211 (.149 \textit{to} .292) ** & 1.602 (1.455 \textit{to} 1.756) **& .407 (.324 \textit{to} .495) ** & 2.309 (2.154 \textit{to} 2.472) ** & \textbf{.667 (.579 \textit{to} .744)} & 2.943 (2.772 \textit{to} 3.114) & .618 (.530 \textit{to} .699) & \textbf{3.146 (2.951 \textit{to} 3.333)}\\
OCD & .183 (.117 \textit{to} .273) ** & 1.828 (1.634 \textit{to} 2.032) ** & .204 (.135 \textit{to} .297) ** & 2.140 (1.989 \textit{to} 2.280) ** & .699 (.599 \textit{to} .783) & 2.968 (2.763 \textit{to} 3.151) & \textbf{.710 (.611 \textit{to} .792)} & \textbf{3.065 (2.828 \textit{to} 3.280)} \\
FBC & .097 (.048 \textit{to} .187) ** & 1.431 (1.278 \textit{to} 1.583) ** & .236 (.153 \textit{to} .346) ** & 2.097 (1.917 \textit{to} 2.278) ** & .597 (.482 \textit{to} .703) & 2.875 (2.653 \textit{to} 3.083) & \textbf{.889 (.796 \textit{to} .943)} & \textbf{3.597 (3.444 \textit{to} 3.750)}\\
Macular & .305 (.235 \textit{to} .385) ** & 1.766 (1.610 \textit{to} 1.929) ** & .532 (.450 \textit{to} .612) ** & 2.369 (2.213 \textit{to} 2.511) ** & \textbf{.809 (.736 \textit{to} .865)} & \textbf{3.014 (2.851 \textit{to} 3.163)} & .624 (.542 \textit{to} .700) ** & 2.851 (2.660 \textit{to} 3.028)\\ 
Vascular & .125 (.043 \textit{to} .310) ** & 1.292 (1.083 \textit{to} 1.542) ** & .375 (.212 \textit{to} .573) & 2.208 (1.917 \textit{to} 2.542) * & .583 (.388 \textit{to} .755) & 2.833 (2.458 \textit{to} 3.208) & \textbf{.875 (.690 \textit{to} .957)} & \textbf{3.667 (3.500 \textit{to} 3.833)}\\
\bottomrule
\end{tabular}
}
\label{Fig3-2}
\vspace{-5mm}
\end{table*}

\vspace{-3mm}
\subsection{\textcolor{Revision}{Comprehensive Evaluation using SBERT Similarity}}
\textcolor{Revision}{To establish performance validation through the comparative analysis, we conduct systematic evaluation against comprehensive baseline architectures encompassing both closed-source API-based systems (Gemini Pro, GPT-4V) and open-source fine-tuned models (InternVL, LLaVA, Qwen-VL, Med-Flamingo, LLaVA-Med, InstructBLIP, Mini-Gemini). All open-source baseline methods undergo identical pretraining and fine-tuning protocols using same pretraining datasets and our MMFundus dataset. Table \ref{ExtendedR2} presents comprehensive empirical evidence establishing VisionUnite's superiority across all other methods. Results demonstrate substantial performance improvements ranging from 5.68\% (vs. LLaVA) to 18.54\% (vs. InstructBLIP), with VisionUnite achieving 79.94\% overall performance. These results conclusively demonstrate that VisionUnite's innovations with generic vision-language approaches, targeted clinical knowledge integration and specialized multimodal reasoning capabilities.}
\vspace{-3mm}
\subsection{\textcolor{Revision}{Multiple-choice Evaluation}}
\textcolor{Revision}{We conduct comprehensive multiple-choice evaluation across 25 ophthalmological conditions, and the results are shown in Table \ref{ExtendedFig5}. Compared with other fine-tuned SOTA methods including LLaVA-Med~\cite{li2024llava}, our proposed VisionUnite achieves superior performance across most conditions, including common diseases such as Age-Related Macular Degeneration (AMD), Diabetic Retinopathy (DR), Glaucoma, and Cataract, as well as underrepresented conditions including Central Serous Retinopathy (CSR), and Tessellation. VisionUnite demonstrates exceptional diagnostic accuracy with an overall performance of 94.36\%, substantially outperforming both closed-source API-based methods (GPT-4V: 75.28\%, Gemini Pro: 69.19\%) and open-source fine-tuned approaches (best performing LLaVA: 85.22\%). The performance advantage underscores the robustness and effectiveness of VisionUnite in accurately diagnosing different ocular conditions represented in clinical practice. We believe that a key factor contributing to the superior performance is its ability to embed explicit knowledge from sign labels, which represent specific clinical features associated with each disease, allowing the model to make more informed and accurate predictions.}
\vspace{-4mm}
\subsection{Diagnosis of Specific Disease and Sign}
Building upon the foundational analysis of average diagnostic performance, our investigation further explores the nuanced diagnostic capabilities of the model and junior ophthalmologist for both healthy ocular conditions and a spectrum of specific ophthalmological pathologies. This segment of our study was particularly comprehensive, encompassing a wide array of conditions ranging from commonplace to rare. Specifically, our focus extended to seven distinct conditions including Age-Related Macular Degeneration (AMD), Branch Retinal Vein Occlusion (BRVO), Cataract, Diabetic Retinopathy (DR), Drusen, Glaucoma, and Myopia as shown in Table \ref{Fig3}. In addition, we explore the diagnostic performance of different signs in Table \ref{Fig3-2}. VisionUnite and ophthalmologists perform significantly better than Gemini Pro and GPT-4V in six common and underrepresented ophthalmic Diseases except for Cataracts. The performance of GPT-4V on cataracts is comparable with VisionUnite and ophthalmologists. Among the Diseases, VisionUnite has the best diagnostic performance in four Diseases: AMD, DR, Drusen, and Glaucoma. For AMD and DR, their performance is not as good as VisionUnite due to the tendency of ophthalmologists to provide incorrect disease staging.
Regarding diagnosis involving different signs, VisionUnite is also better than GPT-4V and Gemini Pro.

\begin{table*}[!t]
\centering
\LARGE
\caption{The joint diagnosis analysis of multiple Diseases. * means p-value $<$ 0.1 and ** means p-value $<$ 0.001.}
\vspace{-2.5mm}
\resizebox{\textwidth}{!}{
\begin{tabular}{l|c|c|c|c|c|c|c|r}
\toprule
\multirow{2}{*}{Method}& \multicolumn{2}{c|}{Gemini Pro~\cite{teamgemini}} & \multicolumn{2}{c|}{GPT-4V~\cite{2023GPT4VisionSC}} & \multicolumn{2}{c|}{VisionUnite} & \multicolumn{2}{c}{Doctor} \\
& \textit{Accuracy (95\% CI)} &\textit{Relevance (95\% CI)}& \textit{Accuracy (95\% CI)} &\textit{Relevance (95\% CI)} & \textit{Accuracy (95\% CI)} &\textit{Relevance (95\% CI)}& \textit{Accuracy (95\% CI)} &\textit{Relevance (95\% CI)}\\
\midrule
Round 1 & .000 (.000 \textit{to} .299) * & 1.222 (1.000 \textit{to} 1.556) ** & .222 (.063 \textit{to} .547) & 2.111 (1.667 \textit{to} 2.556) * & \textbf{.444 (.189 \textit{to} .733)} & \textbf{3.333 (2.889 \textit{to} 3.778)} & .444 (.189 \textit{to} .733) & 3.333 (2.778 \textit{to} 3.778)\\
Round 2 & .111 (.020 \textit{to} .435) ** & 1.778 (1.000 \textit{to} 2.667) * & .333 (.121 \textit{to} .646) * & 2.111 (1.556 \textit{to} 2.667) * & \textbf{.778 (.453 \textit{to} .937)} & 3.000 (2.556 \textit{to} 3.444) & .667 (.354 \textit{to} .879) & \textbf{3.111 (2.444 \textit{to} 3.778)}\\
Round 3 & .222 (.063 \textit{to} .547) * & 1.778 (1.333	\textit{to} 2.333) * & .333 (.121 \textit{to} .646) & 2.556 (2.000 \textit{to} 3.111) & .556 (.267 \textit{to} .811) & 2.778 (2.111 \textit{to} 3.333) & \textbf{.667 (.354 \textit{to} .879)} & \textbf{2.889 (1.889 \textit{to} 3.667)}\\
\midrule
Mean $\uparrow$ & .111 (.039 \textit{to} .281) ** & 1.593 (1.259 \textit{to} 1.963) ** & .296 (.159 \textit{to} .485) * & 2.259 (1.926 \textit{to} 2.593) * & \textbf{.593 (.407 \textit{to} .755)} & 3.037 (2.704 \textit{to} 3.333) & .593 (.407 \textit{to} .755) & \textbf{3.111 (2.667 \textit{to} 3.519)} \\
\bottomrule
\end{tabular}
}
\label{ExtendedFig1}
\vspace{-3.5mm}
\end{table*}

\begin{table*}[!t]
\centering
\LARGE
\caption{The joint diagnosis analysis of multiple Diseases. * means p-value $<$ 0.1 and ** means p-value $<$ 0.001.}
\vspace{-2.5mm}
\resizebox{\textwidth}{!}{
\begin{tabular}{l|c|c|c|c|c|c|c|r}
\toprule
\multirow{2}{*}{Method}& \multicolumn{2}{c|}{Gemini Pro~\cite{teamgemini}} & \multicolumn{2}{c|}{GPT-4V~\cite{2023GPT4VisionSC}} & \multicolumn{2}{c|}{VisionUnite} & \multicolumn{2}{c}{Doctor} \\
& \textit{Accuracy (95\% CI)} &\textit{Relevance (95\% CI)}& \textit{Accuracy (95\% CI)} &\textit{Relevance (95\% CI)} & \textit{Accuracy (95\% CI)} &\textit{Relevance (95\% CI)}& \textit{Accuracy (95\% CI)} &\textit{Relevance (95\% CI)}\\
\midrule
Round 1 & .000 (.000 \textit{to} .194) * & 1.125 (1.000 \textit{to} 1.313) ** & .250 (.102 \textit{to} .495) & 2.188 (1.875 \textit{to} 2.500) ** & .438 (.231 \textit{to} .668) & 3.063 (2.750 \textit{to} 3.375) & \textbf{.688 (.444 \textit{to} .858)} & \textbf{3.625 (3.250 \textit{to} 3.938)}\\
Round 2 & .125 (.035 \textit{to} .360) ** & 1.625 (1.188 \textit{to} 2.188) * & .313 (.142 \textit{to} .556) * & 2.250 (1.813 \textit{to} 2.688) & .688 (.444 \textit{to} .858) & 2.813 (2.313 \textit{to} 3.313) & \textbf{.813 (.570 \textit{to} .934)} & \textbf{3.313 (2.813 \textit{to} 3.688)}\\
Round 3 & .250 (.102 \textit{to} .495) * & 1.563 (1.250 \textit{to} 1.875) ** & .500 (.280 \textit{to} .720) & 2.375 (1.938 \textit{to} 2.813) & .625 (.386 \textit{to} .815) & 2.875 (2.375 \textit{to} 3.375) & \textbf{.750 (.505 \textit{to} .898)} & \textbf{3.188 (2.563 \textit{to} 3.688)}\\
\midrule
Mean $\uparrow$ & .125 (.059 \textit{to} .247) ** & 1.438 (1.250 \textit{to} 1.667) ** & .354 (.234 \textit{to} .496) * & 2.271 (2.063 \textit{to} 2.500) ** & .583 (.443 \textit{to} .712) & 2.917 (2.667 \textit{to} 3.167) & \textbf{.750 (.612 \textit{to} .851)} & \textbf{3.375 (3.104 \textit{to} 3.604)}\\
\bottomrule
\end{tabular}
}
\label{ExtendedFig1-2}
\vspace{-3.5mm}
\end{table*}

\begin{table*}[!t]
\centering
\LARGE
\caption{The misdiagnosis analysis in healthy conditions. * means p-value $<$ 0.1 and ** means p-value $<$ 0.001.}
\vspace{-2.5mm}
\resizebox{\textwidth}{!}{
\begin{tabular}{l|c|c|c|c|c|c|c|r}
\toprule
\multirow{2}{*}{Method}& \multicolumn{2}{c|}{Gemini Pro~\cite{teamgemini}} & \multicolumn{2}{c|}{GPT-4V~\cite{2023GPT4VisionSC}} & \multicolumn{2}{c|}{VisionUnite} & \multicolumn{2}{c}{Doctor} \\
& \textit{Accuracy (95\% CI)} &\textit{Relevance (95\% CI)}& \textit{Accuracy (95\% CI)} &\textit{Relevance (95\% CI)} & \textit{Accuracy (95\% CI)} &\textit{Relevance (95\% CI)}& \textit{Accuracy (95\% CI)} &\textit{Relevance (95\% CI)}\\
\midrule
Round 1 & .519 (.387 \textit{to} .649) ** & 1.827 (1.558 \textit{to} 2.135) ** & .827 (.703 \textit{to} .906) * & 2.346 (2.058 \textit{to} 2.635) ** & \textbf{.962 (.870 \textit{to} .989)} & \textbf{3.269 (3.058 \textit{to} 3.481)} & .846 (.725 \textit{to} .920) * & 2.558 (2.288 \textit{to} 
 2.846) **\\
Round 2 & .519 (.387 \textit{to} .649) ** & 2.135 (1.827 \textit{to} 2.462) * & .731 (.597 \textit{to} .832) * & 2.577 (2.308 \textit{to} 2.846) & \textbf{.904 (.794 \textit{to} .958)} & \textbf{2.750 (2.462 \textit{to} 3.058)} & .885 (.770 \textit{to} .946) & 2.538 (2.269 \textit{to} 2.808)\\
Round 3 & .673 (.538 \textit{to} .785) * & 2.269 (2.000 \textit{to} 2.558) & .846 (.725 \textit{to} .920) & \textbf{2.712 (2.423 \textit{to} 3.019)} & .885 (.770 \textit{to} .946) & 2.558 (2.269 \textit{to} 2.865) & \textbf{.923 (.818 \textit{to} .970)} & 2.462 (2.115 \textit{to} 2.788)\\
\midrule
Mean $\uparrow$ & .571 (.492 \textit{to} .646) ** & 2.077 (1.904 \textit{to} 2.263) ** & .801 (.732 \textit{to} .856) * & 2.545 (2.385 \textit{to} 2.712) * & \textbf{.917 (.863 \textit{to} .951)} &  \textbf{2.859 (2.692 \textit{to} 3.019)} & .885 (.825 \textit{to} .926) & 2.519 (2.359 \textit{to} 2.692) *\\
\bottomrule
\end{tabular}
}
\label{ExtendedFig2}
\vspace{-3.5mm}
\end{table*}

\begin{table*}[!t]
\centering
\LARGE
\caption{The diagnostic continuity analysis in patient interaction includes incorrect prediction (IN.) and correct prediction (CORR.). \\ * means p-value $<$ 0.1 and ** means p-value $<$ 0.001.}
\vspace{-2.5mm}
\resizebox{\textwidth}{!}{
\begin{tabular}{l|c|c|c|c|c|c|c|r}
\toprule
\multirow{2}{*}{Method}& \multicolumn{2}{c|}{Gemini Pro~\cite{teamgemini}} & \multicolumn{2}{c|}{GPT-4V~\cite{2023GPT4VisionSC}} & \multicolumn{2}{c|}{VisionUnite} & \multicolumn{2}{c}{Doctor} \\
& \textit{In. Relevance (95\% CI)} &\textit{Corr. Relevance (95\% CI)}& \textit{In. Relevance (95\% CI)} &\textit{Corr. Relevance (95\% CI)}& \textit{In. Relevance (95\% CI)} &\textit{Corr. Relevance (95\% CI)}& \textit{In. Relevance (95\% CI)} &\textit{Corr. Relevance (95\% CI)}\\
\midrule
Round 1 & 1.317 (1.221 \textit{to} 1.421) ** & 2.629 (2.286 \textit{to} 2.943) ** & 1.879 (1.748 \textit{to} 2.000) ** & 2.671 (2.466 \textit{to} 2.890) ** & \textbf{2.400 (2.156 \textit{to} 2.644)} & 3.430 (3.319 \textit{to} 3.533) & 2.241 (2.000 \textit{to} 2.448) & \textbf{3.443 (3.295 \textit{to} 3.574)} \\
Round 2 & 1.423 (1.300 \textit{to} 1.538) ** & 2.880 (2.620 \textit{to} 3.160) * & 1.830 (1.710 \textit{to} 1.950) & 2.888 (2.688 \textit{to} 3.088) * & \textbf{2.081 (1.784 \textit{to} 2.351)} & 3.189 (3.042 \textit{to} 3.350) & 1.864 (1.614 \textit{to} 2.136) & \textbf{3.250 (3.103 \textit{to} 3.412)}\\
Round 3 & 1.404 (1.270 \textit{to} 1.539) * & 2.659 (2.451 \textit{to} 2.879) * & \textbf{1.875 (1.703 \textit{to} 2.063)} & 2.853 (2.707 \textit{to} 3.017) & 1.703 (1.486 \textit{to} 1.919) & 2.930 (2.755 \textit{to} 3.105) & 1.634 (1.366 \textit{to} 1.927) & \textbf{3.115 (2.942 \textit{to} 3.273)}\\
\midrule
Mean $\uparrow$ & 1.376	(1.310 \textit{to} 1.448) ** & 2.716 (2.568 \textit{to} 2.881) ** & 1.860 (1.779 \textit{to} 1.941) * & 2.814 (2.706 \textit{to} 2.918) ** & \textbf{2.084 (1.924 \textit{to} 2.244)} & 3.178 (3.102 \textit{to} 3.266) & 1.951 (1.797 \textit{to} 2.105) & \textbf{3.262 (3.166 \textit{to} 3.358)} \\
\bottomrule
\end{tabular}
}
\label{ExtendedFig4}
\vspace{-3.5mm}
\end{table*}

\vspace{-4mm}
\subsection{Joint Diagnosis of Multiple Diseases and Signs}
\vspace{-1mm}
When dealing with fundus images from patients suffering from multiple diseases, the overlapping or combined manifestations of these conditions can significantly complicate the diagnostic process~\cite{Kern2015}. Each disease may present with specific signs on fundus examination, such as hemorrhages, exudates, or abnormalities in the vascular architecture. However, when multiple pathologies coexist. For instance, diabetic retinopathy and arteriosclerotic Retinopathy both exhibit retinal vessel changes, but their specific impacts on the retina might slightly differ. Diabetic retinopathy typically shows more microaneurysms and hemorrhages, whereas arteriosclerotic retinopathy involves changes in the retinal arterioles. Therefore, we explore the joint diagnostic performance for multiple diseases and multiple signs, including 15 Diseases and 5 signs. As shown in Table \ref{ExtendedFig1}, the overall performance of VisionUnite is superior to GPT-4V and Gemini Pro in the presence of multiple diseases in fundus images. Compared to ophthalmologists, the overall accuracy of VisionUnite is similar, but the diagnostic relevance is lower. In the first two rounds, VisionUnite performs better than junior ophthalmologists, achieving a diagnostic accuracy of 44.4\% and 77.8\% with diagnostic relevance of 3.333 and 3.0. As for the multi-sign diagnosis, VisionUnite also performs better than GPT-4V and Gemini Pro with the results shown in Table \ref{ExtendedFig1-2}.
\vspace{-3mm}
\subsection{Misdiagnosis Analysis in Healthy Conditions}
\vspace{-1mm}
We also investigate the misdiagnosis of various models and ophthalmologists in the face of healthy fundus images. Since healthy samples often outnumber abnormal samples in the real clinical environment, which means it is important to evaluate the performance of the model and ophthalmologists on healthy samples.
As shown in Table \ref{ExtendedFig2}, the overall misdiagnosis rate of VisionUnite is the lowest with only 8.3\%, where the misdiagnosis rate equals $1-diagnostic\ accuracy$ and its overall diagnostic relevance is 2.859. In the first round of diagnosis, VisionUnite performed much better than other models and is also better than ophthalmologists, with a misdiagnosis rate of only 3.8\% less than 15.4\% of ophthalmologists. The results indicate that VisionUnite can extract good representations of healthy fundus images, thereby achieving a lower misdiagnosis rate through superior reasoning ability.
\vspace{-4mm}
\subsection{Diagnostic Continuity in Patient Interaction}
\vspace{-1mm}
To evaluate the proficiency of models in adhering to and interpreting instructions, we quantify the diagnostic relevance of the responses from each group. Our analysis reveals that in comparison to VisionUnite, both GPT-4V and Gemini Pro frequently yield responses that lack significance. We postulate that this phenomenon may be attributed to instances where keyword triggers elicit unduly cautious reactions from the large models, thereby hindering their ability to engage with queries. This observation underscores a prevalent dependency of the large vision-language models on specific problem prompts, in contrast to the more resilient performance exhibited by VisionUnite. The consistency of VisionUnite in handling variously phrased questions with similar underlying meanings, which we term as 'prompt robustness'.
As shown in Table \ref{ExtendedFig4}, we calculate the diagnostic relevance between each model and junior ophthalmologist in both correct and incorrect predictions. The results indicate that whether the prediction is correct or incorrect, VisionUnite's ability to understand problems and follow instructions is comparable to the junior ophthalmologist. In the case of incorrect predictions, the diagnostic relevance of VisionUnite can still reach 2.084, higher than the ophthalmologists' 1.951. VisionUnite and ophthalmologists perform the best in the first round of diagnosis, while their diagnostic relevances slightly decrease in subsequent diagnoses. We believe it is due to the second and third rounds of diagnosis mainly focusing on clinical explanations and medical opinions, which are relatively general.

\begin{table}[!t]
\centering
\LARGE
\caption{
The diagnostic correction analysis in patient interaction. Overall refers to answering correctly in the second or third round. Round 2 refers to answering correctly in the second round and Round 3 refers to answering correctly in the third round. * means p-value $<$ 0.1 and ** means p-value $<$ 0.001.}
\vspace{-2mm}
\resizebox{\linewidth}{!}{
\begin{tabular}{l|c|c|c}
\toprule
\multirow{2}{*}{Method}& {Gemini Pro~\cite{teamgemini}} & {GPT-4V~\cite{2023GPT4VisionSC}} & {VisionUnite} \\
& \textit{Accuracy (95\% CI)} & \textit{Accuracy (95\% CI)} & \textit{Accuracy (95\% CI)} \\
\midrule
Round 2 & .207 (.149 \textit{to} .280) ** & .280 (.204 \textit{to} .372) ** & \textbf{.667 (.521 \textit{to} .786)}\\
Round 3 & .428 (.350 \textit{to} .509) ** & .523 (.430 \textit{to} .616) * & \textbf{.756 (.613 \textit{to} .858)}\\
\midrule
Overall $\uparrow$ &.476 (.396 \textit{to} .557) ** & .626 (.532 \textit{to} .712) ** & \textbf{.867 (.738 \textit{to} .937)} \\
\bottomrule
\end{tabular}
}
\label{ExtendedFig3}
\vspace{-6mm}
\end{table}

\vspace{-4.5mm}
\subsection{Diagnostic Correction in Patient Interaction}
\vspace{-1mm}
We assume that an advanced vision-language model ought to possess the capability for diagnostic rectification. It implies the inherent ability of models to autonomously amend inaccuracies in their initial diagnosis upon receiving additional information during a subsequent patient interaction. In this context, we focus on assessing the diagnostic precision of these models including VisionUnite, GPT-4V, and Gemini Pro. VisionUnite demonstrated superior corrective accuracy, which significantly surpassed GPT-4V and Gemini Pro. The superior performance of VisionUnite not only highlights the potential of domain-specific large models to exhibit heightened sensitivity in problem recognition within the medical field but also underscores their capacity for agile adaptation. Specifically, the ability of VisionUnite to swiftly recalibrate its responses in alignment with nuanced shifts in problem context, which we term as 'problem sensitivity', stands as a testament to its refined diagnostic acumen and adaptability.
As shown in Table \ref{ExtendedFig3}, we calculate the proportion of correct answers in the second or third round for each model in the case of incorrect answers in the first round. The results demonstrate that VisionUnite has an overall correction accuracy of 86.7\%, which is 24\% and 39\% higher than GPT-4V and Gemini Pro respectively with the p-value less than 0.001.

\vspace{-4mm}
\subsection{Diagnostic Errors Analysis between Ophthalmologist and Large Vision-Language Models}
To analyze the outputs of the models thoroughly, we investigate their diagnostic errors and contrast them with assessments from ophthalmologists. We categorize these errors into two types: missed and incorrect errors. Missed errors pertain to incomplete yet correct diagnoses, while incorrect errors include entirely wrong diagnoses and partially correct diagnoses with additional, irrelevant errors. Our goal is to evaluate the completeness and accuracy of responses, noting omissions or irrelevant inclusions. Additionally, we assess error severity, which is categorized as error-free, minor, or major errors based on the impact on clinical judgment and treatment specificity. Minor errors encompass overlooked partial signs not critical to overall judgment or unnecessary diagnoses, as well as generalized treatment recommendations that lack specificity. Major errors include significantly incorrect responses. We also evaluate the potential for physical or mental harm resulting from the answers, grading potential health risks based on severity and likelihood. No harm is considered error-free, mild or moderate harm is considered minor errors, and serious harm or deaths are considered major errors.
As shown in Table \ref{Fig4} and Table \ref{Fig4-2}, we analyze the diagnostic errors in five different situations: overall, single disease, multiple diseases, single sign, and multiple signs. Specifically, VisionUnite and ophthalmologists perform better than Gemini Pro and GPT-4V in various situations. Overall, VisionUnite performs the best in the "incorrect error" dimension, achieving an error-free rate of 78.15\%, which is 4.82\% higher than that of ophthalmologists. Ophthalmologists perform better in the "missed error" dimension, achieving an error-free rate of 71.85\%. We also find that incorrect errors are more associated with major errors than minor errors. Compared to missed errors, incorrect errors often lead to more serious consequences. \textcolor{Revision}{The diagnostic error analysis also provides evidence of VisionUnite's interpretability advantages by comparing the performance between models and doctors.} The details of the classification criteria for diagnostic errors are in the appendix.

\begin{table}[!t]
\LARGE
\centering
\caption{The diagnostic errors analysis (missing error) of single/multiple Diseases and Signs using diagnostic rate.}
\vspace{-2.5mm}
\resizebox{\linewidth}{!}{
\begin{tabular}{cl|c|c|c|c|c}
\toprule
\multicolumn{2}{c|}{Method}                                     & \multicolumn{1}{c|}{Single Disease} & \multicolumn{1}{c|}{Multiple Diseases} & \multicolumn{1}{c|}{Single Sign} & \multicolumn{1}{c|}{Multiple Signs} & \multicolumn{1}{c}{Overall} \\ \midrule
\multicolumn{1}{c|}{\multirow{3}{*}{Gemini Pro~\cite{teamgemini}}}  & Error-free  &    33.53\%   &   11.11\%  & 34.76\% & 8.33\% & 32.41\% \\ 
\multicolumn{1}{c|}{} & Minor &  12.87\%  &    7.41\%   &     12.60\%  &  12.50\%  &  12.59\%    \\
\multicolumn{1}{c|}{} & Major & 53.61\%  &   81.48\%   &  52.64\%  &  79.17\% &    55.00\%     \\ \midrule
\multicolumn{1}{c|}{\multirow{3}{*}{GPT-4V~\cite{2023GPT4VisionSC}}}      & Error-free  &   48.73\%   &    44.44\%   &   49.80\%   &  35.42\%    &     48.52\%    \\ 
\multicolumn{1}{c|}{}   & Minor &   18.71\%   &  11.11\%  &  17.68\%  &      25.00\%     &  18.33\%  \\ 
\multicolumn{1}{c|}{}   & Major &  32.55\%  &  44.44\%  &  32.52\%  &  39.58\%  &   33.15\%    \\ \midrule
\multicolumn{1}{c|}{\multirow{3}{*}{VisionUnite}} & Error-free  &   66.86\%   &   48.15\%   &   66.87\%   &  56.25\%  &  65.93\% \\ 
\multicolumn{1}{c|}{}   & Minor &  19.88\%  &  25.93\%   &  20.53\%   &   16.67\% &   20.19\%     \\ 
\multicolumn{1}{c|}{}   & Major &  13.26\%  &  25.93\%  &   12.60\%   &  27.08\%   &  13.89\%     \\ \midrule
\multicolumn{1}{c|}{\multirow{3}{*}{Doctor}}      & Error-free  &  72.12\% &  66.67\%  &  70.93\%   &     81.25\%     &   71.85\%     \\
\multicolumn{1}{c|}{}  & Minor & 12.87\%   &   7.41\%    &   13.41\%  &    4.17\%  &  12.59\%    \\ 
\multicolumn{1}{c|}{}  & Major & 15.01\%   &  25.93\%    &   15.65\%  &   14.58\%   &    15.56\% \\ \bottomrule
\end{tabular}}
\label{Fig4}
\vspace{-5mm}
\end{table}

\begin{table}[!t]
\LARGE
\centering
\caption{The diagnostic errors analysis (incorrect error) of single/multiple Diseases and Signs using diagnostic rate.}
\vspace{-2.5mm}
\resizebox{\linewidth}{!}{
\begin{tabular}{cl|c|c|c|c|c}
\toprule
\multicolumn{2}{c|}{Method}                                     & \multicolumn{1}{c|}{Single Disease} & \multicolumn{1}{c|}{Multiple Diseases} & \multicolumn{1}{c|}{Single Sign} & \multicolumn{1}{c|}{Multiple Signs} & \multicolumn{1}{c}{Overall} \\ \midrule
\multicolumn{1}{c|}{\multirow{3}{*}{Gemini Pro~\cite{teamgemini}}}  & Error-free  &        34.31\%    &      11.11\%   &  35.37\% &  10.42\% & 33.15\% \\ 
\multicolumn{1}{c|}{}     & Minor &    3.12\%       &   0.00\%   &  2.64\%  &  6.25\% &  2.96\%  \\
\multicolumn{1}{c|}{}     & Major &  62.57\% &     88.89\%   &  61.99\%  &   83.33\%   &   63.89\%    \\ \midrule
\multicolumn{1}{c|}{\multirow{3}{*}{GPT-4V~\cite{2023GPT4VisionSC}}}      & Error-free  &   50.49\%    &   22.22\%  &  51.02\%  &   29.17\%   & 49.07\% \\ 
\multicolumn{1}{c|}{}    & Minor &  5.46\%  &   3.70\%    &   5.08\%    &    8.33\%     &    5.37\%   \\ 
\multicolumn{1}{c|}{}    & Major &  44.05\%   &  74.07\%  &  43.90\%   &   62.50\%    &   45.56\%     \\ \midrule
\multicolumn{1}{c|}{\multirow{3}{*}{VisionUnite}} & Error-free  & 79.14\%   &   59.26\%   &   80.08\%  & 58.33\% &   78.15\% \\ 
\multicolumn{1}{c|}{}   & Minor &  3.51\%  &  11.11\%  &   3.46\%   &     8.33\%  &     3.89\%     \\ 
\multicolumn{1}{c|}{}   & Major & 17.35\%  &  29.63\%   &   16.46\%   &   33.33\%  &  17.96\%  \\ \midrule
\multicolumn{1}{c|}{\multirow{3}{*}{Doctor}}      & Error-free  &   74.07\%   &  59.26\%   &    73.17\%      &    75.00\%       &      73.33\%     \\
\multicolumn{1}{c|}{}    & Minor &   3.51\%   &  11.11\%   &  3.46\%  &   8.33\%    &     3.89\%   \\ 
\multicolumn{1}{c|}{}    & Major &   22.42\%  &   29.63\%   &  23.37\%  &  16.67\%  &  22.78\% \\ \bottomrule
\end{tabular}}
\label{Fig4-2}
\vspace{-6mm}
\end{table}

\begin{figure*}[!t]
\centerline{\includegraphics[width=\linewidth]{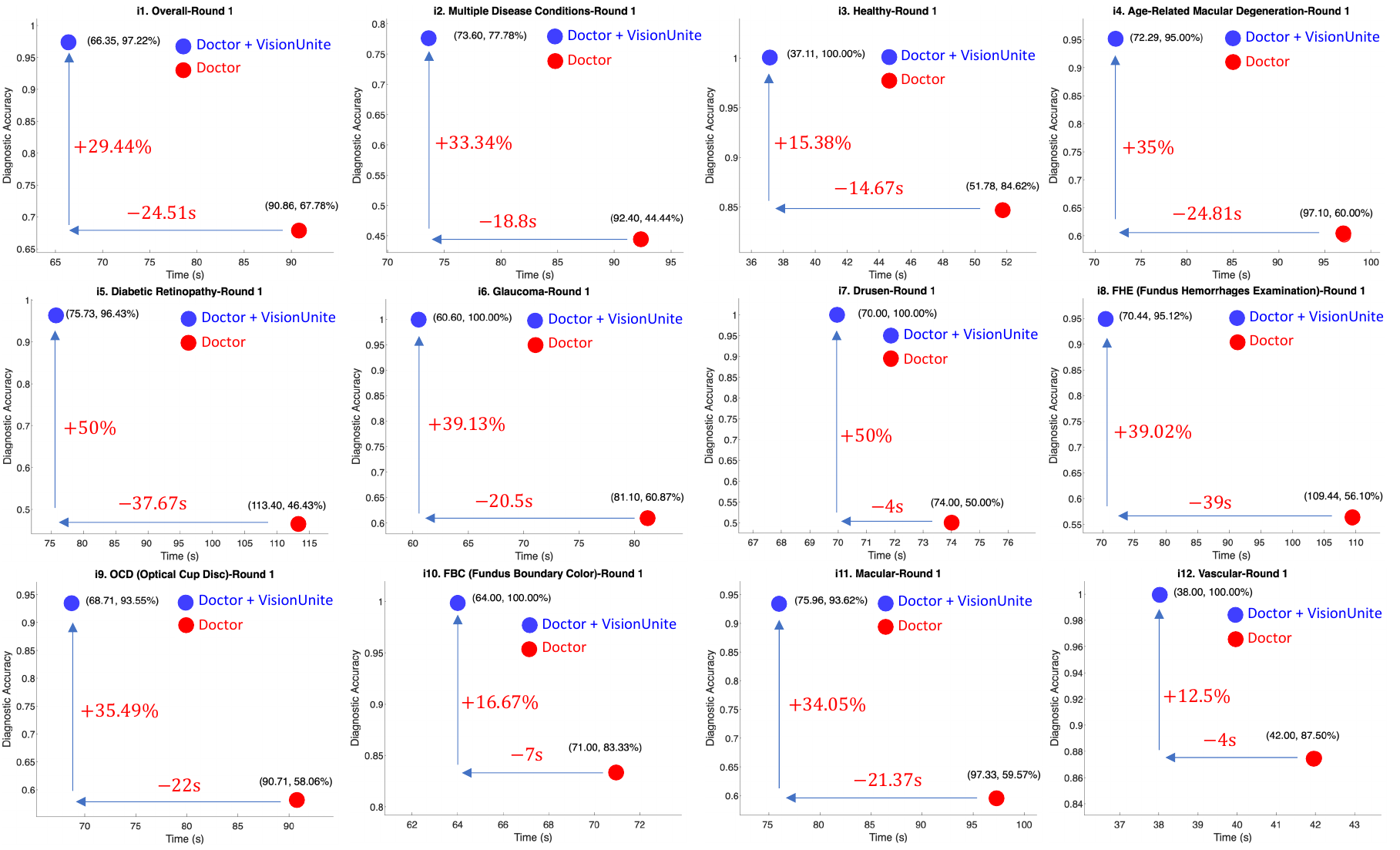}}
\setlength{\abovecaptionskip}{-2mm}
\caption{The diagnostic analysis of VisionUnite assisted junior ophthalmologists (Doctor in the figure). The red dots represent the average diagnostic accuracy and average diagnostic time of the first round only from the Doctor. The blue dots represent the average diagnostic accuracy and average diagnostic time of the first round from Doctor+VisionUnite. The figure contains a diagnostic analysis of the overall situation and the other 11 situations.}
\label{Fig5}
\vspace{-4mm}
\end{figure*}

\begin{figure*}[!t]
\setlength{\abovecaptionskip}{-2mm}
\centerline{\includegraphics[width=\linewidth]{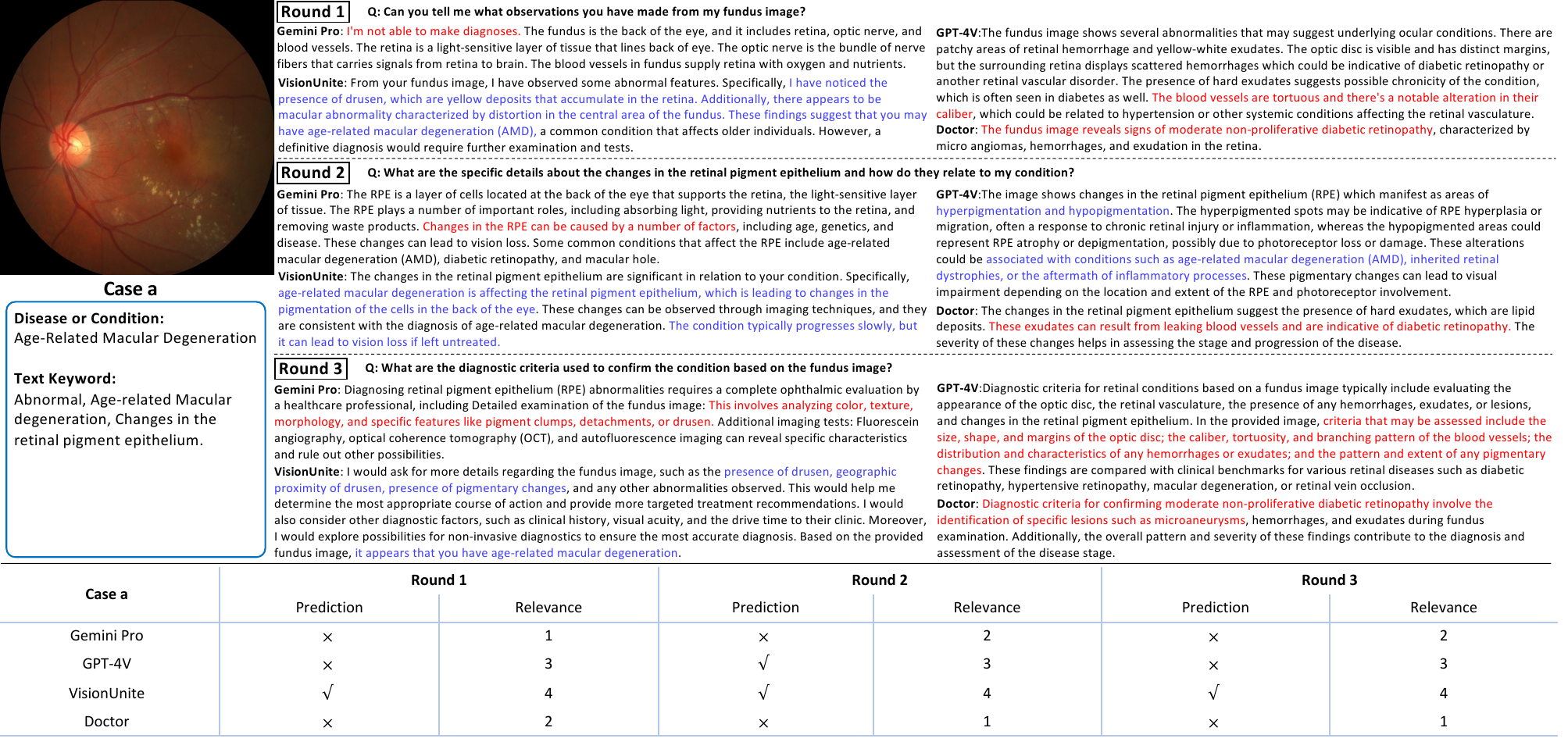}}
\caption{The consistent interpretation analysis of visual and language features between large vision-language models and the junior ophthalmologist (Doctor). \textcolor{red}{Red} represents inaccurate or irrelevant descriptions. \textcolor{blue}{Blue} represents the accurate descriptions. The expert (senior ophthalmologist) evaluation is also provided.}
\label{Fig6}
\vspace{-5mm}
\end{figure*}

\vspace{-3mm}
\subsection{Diagnostic Analysis of VisionUnite Assisted Doctor}
To further explore the clinical decision-support capabilities of VisionUnite, we assess its impact on the diagnostic performance of primary ophthalmologists. We specifically evaluate the improvements in diagnostic accuracy and efficiency when VisionUnite assists ophthalmologists in the initial round which includes 180 questions. Our analysis shows that VisionUnite facilitates a 26.98\% reduction in diagnostic time and a 29.44\% increase in diagnostic accuracy overall illustrated in Figure \ref{Fig5}. Notably, the use of VisionUnite leads to a remarkable 50\% improvement in diagnostic accuracy and a 33\% reduction in diagnostic time in cases of diabetic retinopathy. Furthermore, we examine the impact of VisionUnite on diagnoses involving five distinct physical signs detailed in Figure \ref{Fig5} i8-i12. The assistance is most significant for fundus hemorrhage exudates (FHE) and optic disc cupping (OCD), where VisionUnite helps improve diagnostic accuracy by 39.02\% and 35.49\%, respectively, while also reducing diagnostic time by 35.63\% and 24.25\%. In traditional fundus lesion screening, the reliance on the subjective judgment of ophthalmologists based on fundus photography can lead to variability in diagnosis owing to human factors such as fatigue, variability in training, and personal interpretative skills. This subjectivity can contribute to higher risks of missed diagnoses and misdiagnoses. In contrast, VisionUnite can standardize the interpretation of fundus images by consistently applying the same criteria to analyze and interpret data across all instances. In diseases with multifaceted presentations, such as diabetic retinopathy, VisionUnite can analyze multiple aspects of the disease, providing a holistic view that aids in a more comprehensive assessment. Additionally, predictive diagnostics further assist doctors in developing and evaluating treatment strategies. 

\vspace{-4mm}
\subsection{Consistent Interpretation of Visual and Language Features}
We explore the coherent integration of visual and textual elements within large vision-language models, an endeavor to assess the alignment between the textual descriptions and the corresponding visual data. Our investigation addresses the potential for 'illusory discrepancies' within the output of these sophisticated models. Specifically, we list the outputs of VisionUnite against those from other large vision-language models, with a keen focus on discerning any textual outputs that may depict features not present within the associated imagery. For instance, in cases where the diagnosis pertained to age-related macular degeneration, and the imagery exclusively showcases the presence of drusen, we should evaluate whether the textual narrative unjustifiably extended the diagnosis to encompass additional manifestations such as hemorrhaging and exudation as shown in Figure \ref{Fig6}. Such a detailed examination is pivotal for understanding and mitigating the tendency of these models to over-interpret or misalign textual descriptions with their visual counterparts, thereby ensuring a more accurate and reliable diagnostic output. Meanwhile, we also present a series of evaluation criteria for the responses of the models and junior ophthalmologists in this section.

\vspace{-3mm}
\subsection{\textcolor{Revision}{Ablation Studies of Proposed Components}}
\textcolor{Revision}{To validate individual contribution of our proposed component, we conduct ablation studies across our multi-round VQA evaluation framework. Table \ref{ExtendedR1} presents comprehensive results demonstrating the necessity and efficacy of each component. The results demonstrate that each proposed component contributes meaningfully to diagnostic performance.  Individual component analysis demonstrates that the Vision Adapter with classification loss ($L_{CLS}$) achieves the highest standalone performance (75.95\%), followed closely by the contrastive learning component ($L_{CLIP}$) at 75.14\%, while the Vision Projector yields 74.63\% when used in isolation. This performance hierarchy indicates that sign-level feature extraction and image-text alignment represent the most fundamental capabilities for ophthalmological diagnosis, with the Vision Adapter's explicit clinical knowledge encoding providing slight advantages over general visual-linguistic alignment. The relatively lower performance of the Vision Projector in isolation suggests its primary role as a facilitating component that enhances the effectiveness of other modules rather than serving as a standalone diagnostic feature extractor.}
\vspace{-4mm}
\begin{table}[!t]
\centering
\LARGE
\caption{\textcolor{Revision}{The ablation study of each proposed component on multi-rounds VQA dataset using SBERT Similarity (\%). Due to the dependence of $L_{CLS}$ on Vision Adapter, Vision Adapter and $L_{CLS}$ are considered as the same component.}}
\vspace{-2.5mm}
\resizebox{\linewidth}{!}{
\begin{tabular}{ccc|ccc|c}
\toprule
$L_{CLIP}$& Vision Adapter($L_{CLS}$) & Vision Projector & Round 1 & Round 2 & Round 3 & Overall $\uparrow$ \\
\midrule
\checkmark &  &  & 76.83 & 75.04 & 73.55 & 75.14 \\
 & \checkmark &  & 77.62 & 75.37 & 74.86 & 75.95 \\
 &  & \checkmark & 76.41 & 74.28 & 73.19 & 74.63 \\
\checkmark & \checkmark &  & 82.45 & 77.81 & 76.93 & 79.06 \\
\checkmark &  & \checkmark & 79.35 & 76.79 & 75.91 & 77.35 \\
 & \checkmark & \checkmark & 81.42 & 77.15 & 76.34 & 78.30 \\
\checkmark & \checkmark & \checkmark & \textbf{83.46} & \textbf{78.53} & \textbf{77.82} & \textbf{79.94}\\
\bottomrule
\end{tabular}
}
\label{ExtendedR1}
\vspace{-4mm}
\end{table}

\vspace{-2mm}
\subsection{\textcolor{Revision}{Ablation Studies of Pretraining and/or Finetuning Datasets}}
\textcolor{Revision}{To validate the individual contributions of pretraining and fine-tuning datasets and ensure fair comparison with baseline configurations, we conduct systematic ablation studies across multiple dataset combinations. This analysis isolates the performance impact of each dataset component, providing transparent evaluation of our methodological contributions. We evaluate six distinct configurations using our established multi-round VQA evaluation protocol with SBERT similarity assessment, ranging from a baseline configuration using only MMFundus fine-tuning to our complete configuration incorporating all pretraining datasets. Table \ref{ExtendedR3} presents comprehensive results demonstrating systematic performance improvements across dataset configurations, with the baseline configuration achieving 74.69\% overall performance and establishing a clear comparative foundation. The results reveal that MS COCO pretraining contributes a 2.96\% improvement (77.65\% vs 74.69\%), demonstrating the value of general vision-language representations for ophthalmological applications, while PMC-OA pretraining provides a 2.07\% improvement (76.76\% vs 74.69\%), validating the importance of biomedical domain knowledge integration. The combined pretraining datasets yield a 4.69\% improvement (79.38\% vs 74.69\%), indicating complementary benefits from diverse pretraining sources and confirming the synergistic effects of our comprehensive training approach. It should be noted that we specifically modified the format and style of RET. annotations to ensure that VisionUnite could generate responses in the same style without finetuning on MMFundus. The results demonstrate consistent performance patterns across all three dialogue rounds, with the complete configuration achieving optimal performance in Round 1 (83.46\%), Round 2 (78.53\%), and Round 3 (77.82\%).}
\vspace{-4mm}
\begin{table}[!t]
\LARGE
\centering
\caption{\textcolor{Revision}{The ablation study of pretraining and/or finetuning datasets on multi-rounds VQA dataset using SBERT Similarity (\%). RET. indicates Retina Image Bank dataset.}}
\vspace{-2mm}
\resizebox{\linewidth}{!}{
\begin{tabular}{ccc|c|ccc|c}
\toprule
\multicolumn{3}{c|}{\textbf{Pretraining datasets}}  & \multicolumn{1}{c|}{\textbf{Finetuning dataset}} & \multirow{2}{*}{Round 1} & \multirow{2}{*}{Round 2} & \multirow{2}{*}{ Round 3} & \multirow{2}{*}{Overall $\uparrow$}\\
MS COCO & PMC-OA & RET. & MMFundus & & & &\\
\midrule
 & & & \checkmark & 76.35 & 74.58 & 73.14 & 74.69 \\
\checkmark &  &  & \checkmark & 80.92 & 76.44 & 75.59 & 77.65 \\
 & \checkmark &  & \checkmark & 79.27 & 76.05 & 74.96 & 76.76 \\
\checkmark & \checkmark & & \checkmark & 82.78 & 78.12 & 77.24 & 79.38\\
\checkmark & \checkmark & \checkmark &  & 75.51 & 73.83 & 72.33 & 73.89 \\
\checkmark & \checkmark & \checkmark & \checkmark & \textbf{83.46} & \textbf{78.53} & \textbf{77.82} & \textbf{79.94}\\
\bottomrule
\end{tabular}
}
\label{ExtendedR3}
\vspace{-6mm}
\end{table}

\section{Discussion}
The diagnostic efficacy of VisionUnite in ophthalmic conditions has been extensively validated in various clinical settings, indicating improvements in initial screenings, thereby enhancing healthcare efficiency, especially in under-resourced areas. The image analysis and diagnostic capabilities of VisionUnite are comparable to junior ophthalmologists, sometimes even surpassing them. It outperforms other large vision-language models like GPT-4V and Gemini Pro by providing more accurate diagnoses and clearer explanations, crucial for clinical use. VisionUnite also accelerates disease diagnosis, supports patient interaction, and aids in educational efforts for healthcare professionals by providing detailed reports on fundus photography, a feature that traditional models without textual descriptions lack. \textcolor{Revision}{Our model employs a modular design specifically engineered for component-wise optimization, allowing selective fine-tuning of individual components based on computational resources. This approach enables a progressive deployment strategy where resource-constrained environments can implement a reduced version with fewer classification tasks while maintaining core diagnostic functionality. While preliminary evaluations suggest VisionUnite may approximate performance characteristics of junior ophthalmologists under controlled conditions, significant limitations constrain its clinical utility. VisionUnite demonstrates suboptimal performance in several ophthalmic conditions, primarily due to data scarcity within  datasets. Currently restricted to fundus imaging, VisionUnite exhibits the narrow assessment capabilities typical of current AI diagnostic tools, with limited classification labels that restrict detailed diagnostic assessments. Future developments may expand to additional ophthalmic imaging modalities and implement more granular sign-level labeling to enhance diagnostic accuracy.} As technology and data availability improve, VisionUnite is anticipated to handle a wider range of conditions, potentially identifying connections between ocular signs and systemic diseases, thus advancing both ophthalmic and general medical diagnostic processes.
\vspace{-2mm}
\section{Conclusion}
In this study, we propose VisionUnite, which represents a significant advancement as a large vision-language foundation model for ophthalmology with clinical knowledge. Its distinguishing feature lies in its open diagnostic capability for eye diseases, eliminating the requirement for predefined disease ranges and aligning more closely with the demands of clinical diagnosis. Furthermore, the utilization of VisionUnite holds the promise of expediting the identification of previously undiscovered connections between diseases and ocular features, contributing to the refinement of diagnosis.
\vspace{-2mm}
\section*{Acknowledgments}
This research is supported by Shanghai Artificial Intelligence Laboratory. This work is partially supported by the National Key R\&D Program of China (NO.2022ZD0160102, NO.2022YFC2502800), partially supported by the National Natural Science Foundation of China (62272450, 82471066), and partially supported by the China Postdoctoral Science Foundation (Grant No.2022M721743). This work was done during Zihan Li's internship at Shanghai Artificial Intelligence Laboratory.
\vspace{-2mm}

\bibliographystyle{IEEEtran}
\bibliography{main}

\vspace{-60pt}
\begin{IEEEbiography}[{\includegraphics[width=1in,height=1.25in,clip,keepaspectratio]{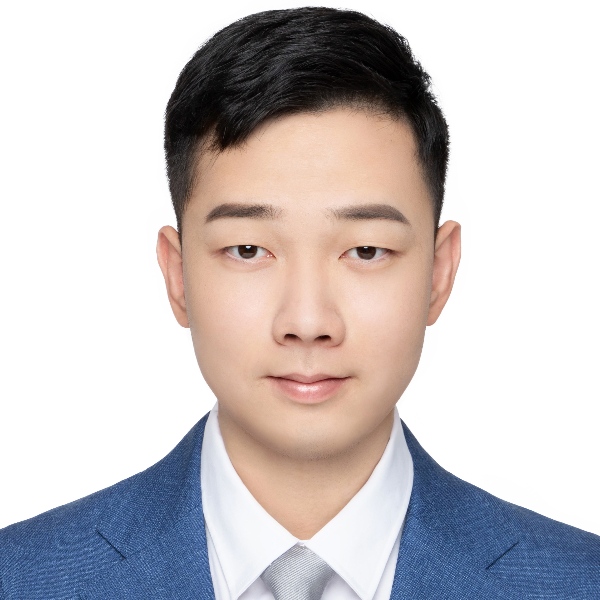}}]{Zihan Li} received the Bachelor’s degree from Xiamen University, Xiamen, China, and the Master’s degree from the University of Illinois at Urbana-Champaign, Champaign, IL, USA. He is currently pursuing the Ph.D. degree with the University of Washington, Seattle, WA, USA. His research interests include computer vision, self-supervised learning, medical image analysis, and vision-language models.
\end{IEEEbiography}

\vspace{-33pt}
\begin{IEEEbiography}[{\includegraphics[width=1in,height=1.25in,clip,keepaspectratio]{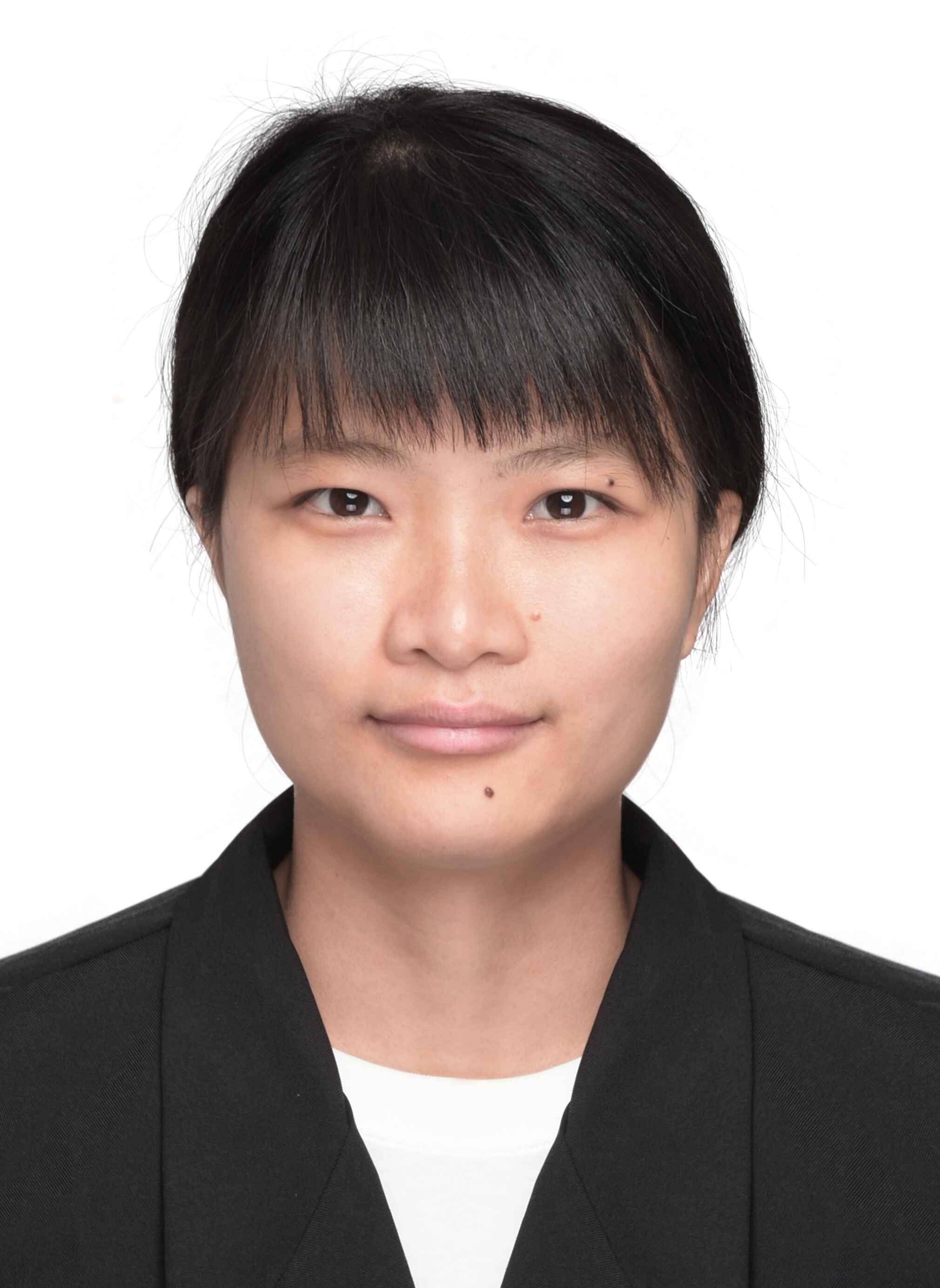}}]{Diping Song} received the PhD degree from the Shenzhen Institute of Advanced Technology at University of Chinese Academy of Sciences in 2022. Now she is an Young Researcher in Shanghai AI Laboratory, China. Her research interests lie on the computer vision, medical image analysis and vision-language models.
\end{IEEEbiography}

\vspace{-33pt}
\begin{IEEEbiography}[{\includegraphics[width=1in,height=1.25in,clip,keepaspectratio]{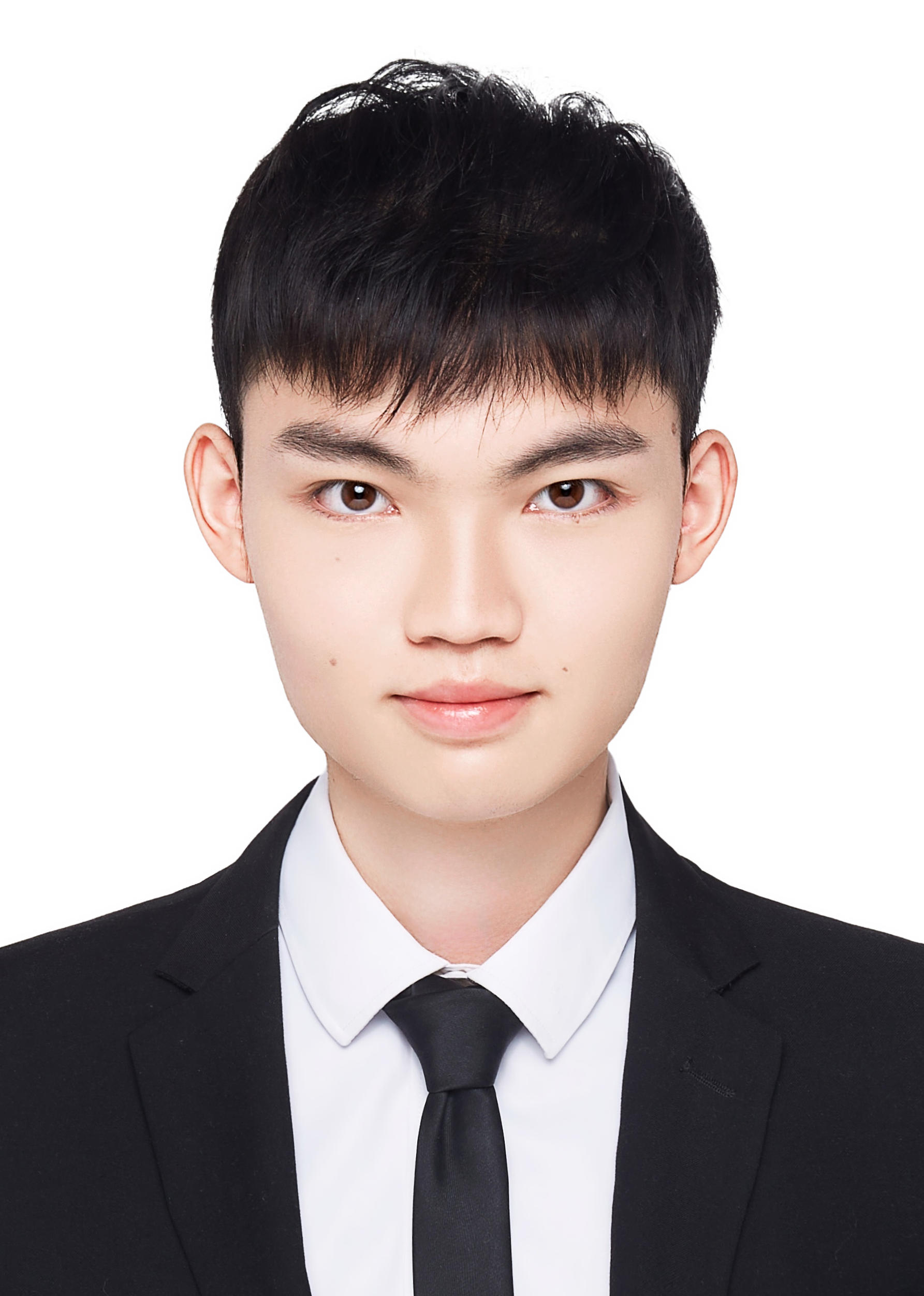}}]{Zefeng Yang} is currently pursuing the Ph.D. degree in Ophthalmology at the Zhongshan Ophthalmic Center, Sun Yat-sen University. He has published two first-author SCI papers in ``Progress in Retinal and Eye Research" and the ``Asia-Pacific Journal of Ophthalmology", and contributed to five additional papers in prestigious journals like Cell Reports Medicine. His main research focus is on the application of artificial intelligence in ophthalmology. 
\end{IEEEbiography}

\vspace{-33pt}
\begin{IEEEbiography}[{\includegraphics[width=1in,height=1.25in,clip,keepaspectratio]{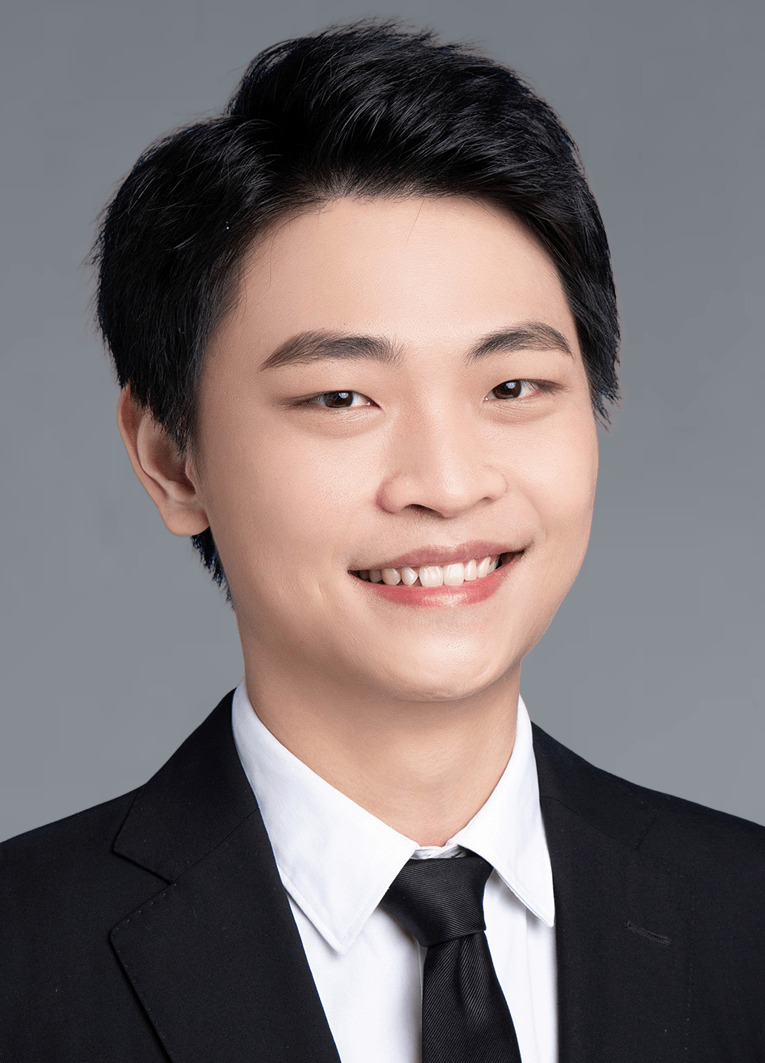}}]{Deming Wang} received the bachelor’s degree from Southern Medical University in 2022 and is currently pursuing a master's degree at the Zhongshan Ophthalmic Center, Sun Yat-sen University. He has contributed to two other top-tier journals, ``Cell Reports Medicine" and ``Ophthalmology". His primary research interests focus on high myopia, glaucoma imaging, and the application of artificial intelligence in ophthalmology.
\end{IEEEbiography}

\vspace{-33pt}
\begin{IEEEbiography}[{\includegraphics[width=1in,height=1.25in,clip,keepaspectratio]{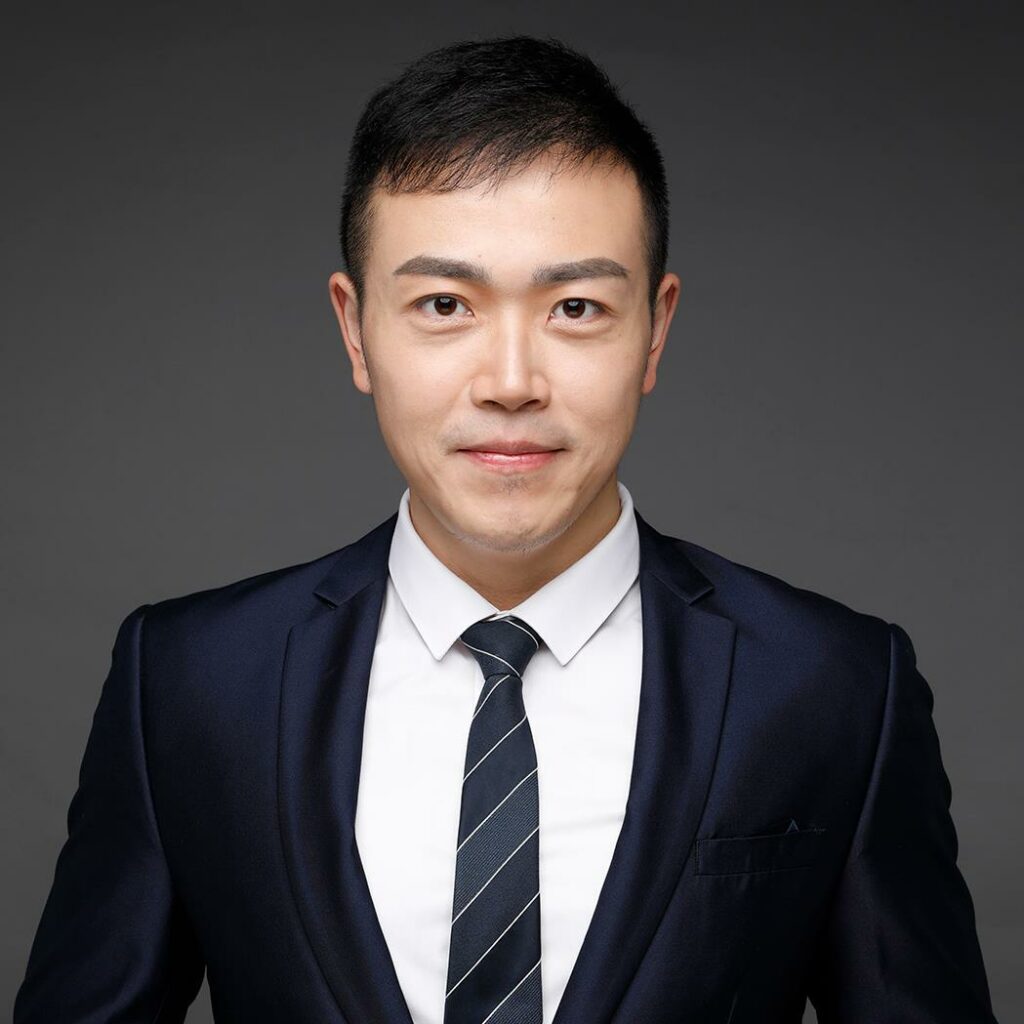}}]{Fei Li} MD, PhD, is an Assistant Researcher at the Zhongshan Ophthalmic Center, dedicated to exploring the forefront of ophthalmic AI. During this period, he has undertaken one National Natural Science Foundation of China (NSFC) Youth Fund project, participated in one NSFC General Program, one sub-project of the Ministry of Science and Technology's key R\&D plan on medical AI. As a co-founder, he has established the open ophthalmic image database iChallenge.
\end{IEEEbiography}

\vspace{-35pt}
\begin{IEEEbiography}[{\includegraphics[width=1in,height=1.25in,clip,keepaspectratio]{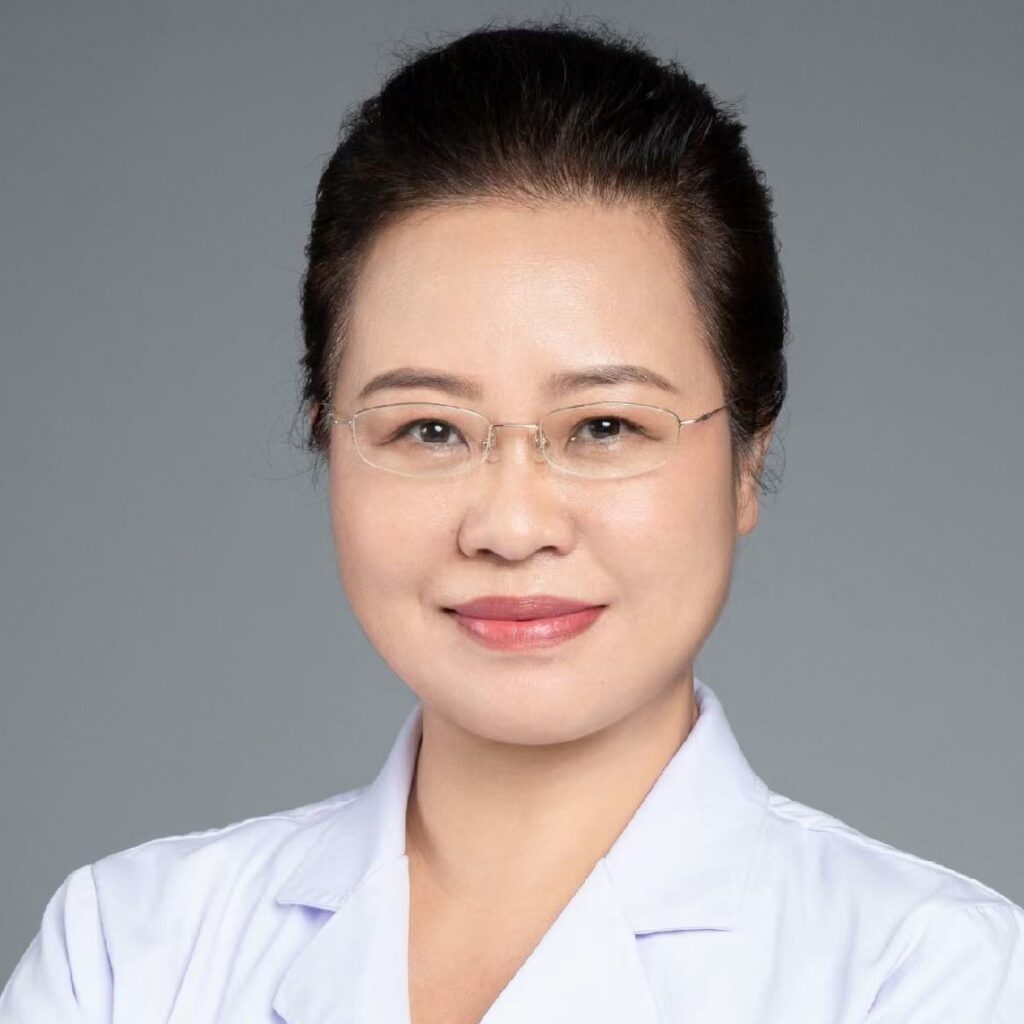}}]{Xiulan Zhang} MD, PhD, glaucoma specialist, is currently at Zhongshan Ophthalmic Center (ZOC), Sun Yat-sen University, P.R. China. She is the outstanding PI of State Key Laboratory of Ophthalmology in China. She is the founder of the first Clinical Research Center for Ophthalmology in China that meets international standards. Prof. Zhang has been deliciated to clinical practice, teaching and research of ophthalmology for 35 years. Prof. Zhang is the pioneer of cutting-edge research on glaucoma AI, imaging and clinical research.
\end{IEEEbiography}

\vspace{-30pt}
\begin{IEEEbiography}[{\includegraphics[width=1in,height=1.25in,clip,keepaspectratio]{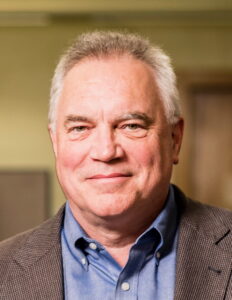}}]{Paul E. Kinahan}(Life Fellow, IEEE) is currently a member of the UW Imaging Research Laboratory. He was a part of the group that built the first prototype combined PET/CT scanner and has also contributed to the current class of data processing image reconstruction algorithms used in PET/CT oncology imaging. He moved to the University of Washington, in 2001, where he continues his research in PET/CT imaging. He has served on committees for RSNA, AAPM, SNM, NIH, and IEEE.
\end{IEEEbiography}

\vspace{-33pt}
\begin{IEEEbiography}[{\includegraphics[width=1in,height=1.25in,clip,keepaspectratio]{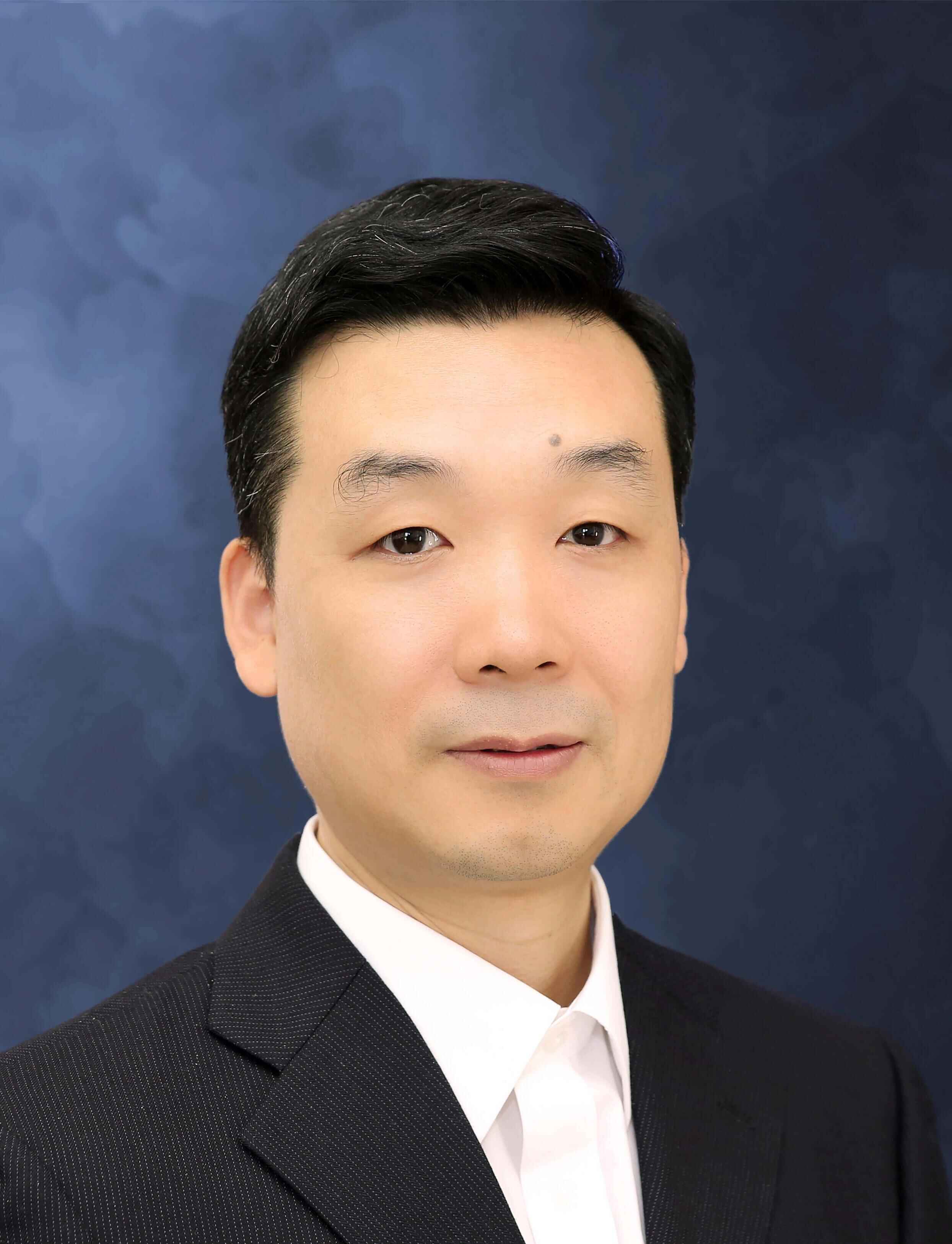}}]{Yu Qiao}(Senior Member, IEEE) is a professor with Shanghai AI Laboratory. His research interests include computer vision, deep learning, and bioinformation. He has published more than 300 papers in international journals and conferences, including T-PAMI, IJCV, T-IP, T-SP, CVPR, ICCV etc. His H-index is over 100, with more than 90,000 citations in Google Scholar. He is a recipient of the distinguished paper award in AAAI 2021. He is an associate editor of Pattern Recognition, Neural Networks, and JVCI. He served as program chair of IEEE ICIST 2014.
\end{IEEEbiography}
\vfill

\end{document}


\title{Appendix of VisionUnite: A Vision-Language Foundation Model for Ophthalmology Enhanced with Clinical Knowledge}
\author{Zihan Li, Diping Song, Zefeng Yang, Deming Wang, Fei Li, Xiulan Zhang,\\ Paul E. Kinahan,~\IEEEmembership{Life Fellow,~IEEE,} Yu Qiao, ~\IEEEmembership{Senior Member,~IEEE}
\thanks{Zihan Li is with Shanghai Artificial Intelligence Laboratory, Shanghai, 200232, China and the University of Washington, Seattle, WA 98195, USA.}
\thanks{Diping Song is with Shanghai Artificial Intelligence Laboratory, Shanghai, 200232, China.}
\thanks{Yu Qiao is with Shanghai Artificial Intelligence Laboratory, Shanghai, 200232, China and Shenzhen Institutes of Advanced Technology, Chinese Academy of Sciences,  Shenzhen, 518055, China.}
\thanks{Zefeng Yang, Deming Wang, Fei Li, and Xiulan Zhang are with State Key Laboratory of Ophthalmology, Zhongshan Ophthalmic Center, Sun Yat-sen University, Guangdong Provincial Key Laboratory of Ophthalmology and Visual Science, Guangdong Provincial Clinical Research Center for Ocular Diseases, Guangzhou, 510060, China}
\thanks{Paul E. Kinahan is with the Department of Bioengineering and the Department of Radiology, University of Washington, Seattle, WA 98195, USA.}
\thanks{Zihan Li and Diping Song have equal contributions to this work.}
\thanks{Corresponding author: Yu Qiao}
}

\markboth{Journal of \LaTeX\ Class Files,~Vol.~18, No.~9, September~2020}%
{How to Use the IEEEtran \LaTeX \ Templates}

\maketitle



\begin{table}[!ht]
\centering
\setlength{\tabcolsep}{6mm}
\caption{Characteristics of patients in the Private dataset.}
\begin{tabular}{ccc}
\toprule
                                                   & \multicolumn{2}{c}{\textbf{Private dataset}}                                        \\
                                                   & \multicolumn{2}{c}{\textbf{(n = 33029)}}                                       \\\cmidrule(l){2-3}
\multirow{-3}{*}{\textbf{Variables}}               & N         &    \%       \\
\midrule
\multicolumn{1}{l}{\textbf{Gender}}                                    &                    &       \\
{Female} & 17735                & 53.70\% \\
{Male}                                               & 15205                & 46.03\% \\
{N/A}                                                & 89                   & 0.27\%  \\
\cmidrule{1-1}
\multicolumn{1}{l}{\textbf{Age (years), Median (IQR)}}& \multicolumn{2}{c}{39.75 (28.0-49.0)} \\
$<$60                                      & 29187                & 88.37\% \\
$>=$60                                                & 3014                 & 9.12\%  \\
N/A                                                & 828                  & 2.51\%  \\
\cmidrule{1-1}
\multicolumn{1}{l}{\textbf{Eye}}                   & \multicolumn{1}{l}{} &       \\
OD                                                 & 16845                & 51.00\% \\
OS                                                 & 16184                & 49.00\% \\
\cmidrule{1-1}
\multicolumn{1}{l}{\textbf{High myopia}}           &                    &       \\
With                                               & 19592                & 59.32\% \\
Without                                            & 13437                & 40.68\% \\
\cmidrule{1-1}
\multicolumn{1}{l}{\textbf{Glaucoma}}              &                    &       \\
With                                               & 5747                 & 17.40\% \\
Suspect                                            & 4859                 & 14.71\% \\
Without                                            & 22423                & 67.89\% \\
\cmidrule{1-1}
\multicolumn{1}{l}{\textbf{High intraocular pressure}} &                    &       \\
With                                               & 372                  & 1.12\%  \\
Suspect                                            & 12                   & 0.04\%  \\
Without                                            & 32645                & 98.84\% \\
\cmidrule{1-1}
\multicolumn{1}{l}{\textbf{Cup-to-Disc ratio increase}} &                    &       \\
With                                               & 1379                 & 4.18\%  \\
Without                                            & 31650                & 95.82\% \\
\cmidrule{1-1}
\multicolumn{1}{l}{\textbf{Cup-to-Disc ratio asymmetry}} &                    &       \\
With                                               & 237                  & 0.72\%  \\
Without                                            & 32792                & 99.28\% \\
\cmidrule{1-1}
\multicolumn{1}{l}{\textbf{Family history}}        &                    &       \\
With                                               & 2272                 & 6.88\%  \\
Without                                            & 30757                & 93.12\%  \\
\bottomrule 
\end{tabular}
\label{tab1}
\vspace{-2mm}
\end{table}

\begin{table}[!ht]
\Large
\setlength{\abovecaptionskip}{0mm}
  \centering
  \caption{\textcolor{Revision}{The label distribution of the multiple-choice evaluation benchmark.}}
  \resizebox{\columnwidth}{!}{%
    \begin{tabular}{c|c|c|c}
    \toprule
    \textbf{Label} & Num & \textbf{Label} & Num \\
    \midrule
    Health& 890 & Other disease & 50\\
    Myopia & 226 & Tessellation & 12\\
    Retinitis & 9 & Chorioretinitis & 3\\
    Diabetic Retinopathy & 149 & Drusen & 30 \\
    Media Haze & 31 & Central Serous Retinopathy & 7 \\
    Cataract & 20 & Arteriosclerotic Retinopathy & 2 \\
    Optic Disk Cupping & 32 & Optic Disc Edema & 11 \\
    Optic Disc Pallor & 2 & Hypertensive Retinopathy & 3\\
    Branch Retinal Vein Occlusion & 16 & Central Retinal Vein Occlusion & 11 \\
    Age-related Macular Degeneration & 171 & No Age-related Macular Degeneration & 311 \\
    Diabetic Macular Edema & 58 & No Macular Edema & 11 \\
    Glaucoma & 162 & No Glaucoma & 12 \\
    \cline{3-4}
    Choroidal Neovascularization & 4 & \textbf{Summary} & \textbf{2233 images}\\
    \bottomrule
    \end{tabular} 
    }
  \label{labeldis}%
  \vspace{-3mm}
\end{table}%

\begin{table*}[!t]
\caption{The classification criteria for diagnostic errors include missing and incorrect errors, as well as minor and major errors. The classification criteria are applied to the analysis of diagnostic errors in this study.}
\vspace{-2mm}
\centerline{\includegraphics[width=\linewidth]{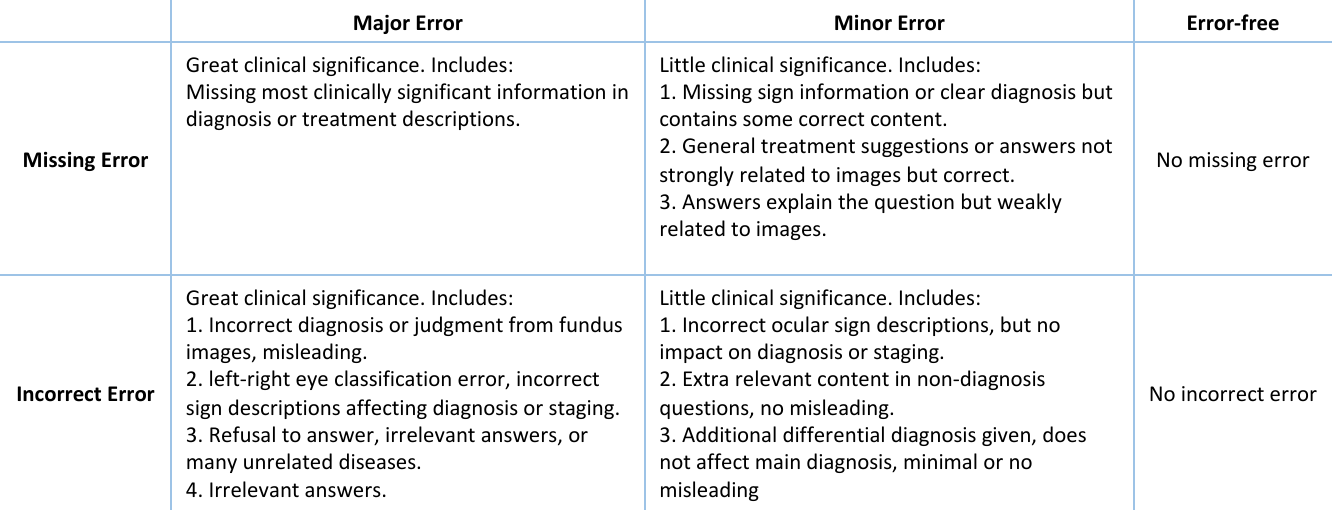}}
\label{tab2}
\vspace{-1mm}
\end{table*}

\subsection{The characteristics of patients in the private dataset} The detailed information of patients in the private dataset is shown in Table \ref{tab1}. The gender ratio and OD/OS ratio are balanced in the dataset.

\subsection{\textcolor{Revision}{The abbreviation in the multiple-choice evaluation}}
\textcolor{Revision}{AMD: Age-related Macular Degeneration; AR: Arteriosclerotic Retinopathy; BRVO: Branch Retinal Vein Occlusion; CN: Choroidal Neovascularization; CRVO: Central Retinal Vein Occlusion; CSR: Central Serous Retinopathy; DR: Diabetic Retinopathy; HR: Hypertensive Retinopathy; DME: Diabetic Macular Edema; MH: Media Haze; ODC: Optic Disk Cupping; ODE: Optic Disc Edema; ODP: Optic Disc Pallor. More details can be seen at Table \ref{labeldis}.}

\subsection{\textcolor{Revision}{Specific version of baseline methods}}
\textcolor{Revision}{Closed-source API-based systems: Gemini Pro (gemini-pro-vision), GPT-4V (gpt-4-1106-vision-preview).}

\textcolor{Revision}{Open-source fine-tuned models: InternVL (InternVL-1.0), LLaVA (LLaVA-1.5), Qwen-VL (Qwen-VL-Chat), Med-Flamingo (Med-Flamingo-V1), LLaVA-Med (LLaVA-Med-V1), InstructBLIP (InstructBLIP-V1), Mini-Gemini(Mini-Gemini-V1).}

\subsection{Construction of the dialogue in the pretrain dataset}
For the question section of the pretrain dataset, we have constructed 20 questions, categorized into two types: indicative long answers (10 sentences) and short answers (10 sentences). Depending on the length of the dialogue answer, a sentence from the relevant question type is selected as the interrogative component of the dialogue. The details are as follows:

\textbf{Short instructions:}
1. Briefly depict the image.
2. Provide a concise overview of the presented image.
3. Summarize the visual elements in a succinct manner.
4. Give a clear, short explanation of the image.
5. Offer a compact interpretation of the provided image.
6. Share a brief account of the key features captured in the photo.
7. Relay a clear and concise description of the shown picture.
8. Render a succinct summary of the photo's content.
9. Craft a compact narrative encapsulating the presented picture.
10. Create a brief, informative summary of the visual content.
\textbf{Long instructions:}
1. Elaborate on the specifics of the given image.
2. Offer an intricate explanation of the visual content.
3. Share a comprehensive rundown of the image presented.
4. Conduct a thorough analysis of the elements within the image.
5. Explain in detail the various aspects portrayed in the image.
6. Characterize the image through a well-detailed description.
7. Analyze the image comprehensively, delving into its details.
8. Illustrate the image through a descriptive explanation.
9. Examine the image closely and articulate its intricate details.
10. Craft an exhaustive depiction of the given image.

\subsection{Data Automatic Generation in MMFundus Dataset}
\textbf{Text Description Automatic Generation:}
We automatically generate text descriptions of the fundus images based on the original label information and generate corresponding sign labels for each image. The text description we design consists of three parts: normal/abnormal, specific diseases or conditions, and clinical explanations related to the disease. For example "Abnormal, Severe Diabetic Macular Edema [specific disease or condition 1], lots of hard exudates near to macula center observed [clinical explanation 1], Moderate Non-Proliferative Diabetic Retinopathy [specific disease or condition 2], Retinal hemorrhages or hard exudates observed [clinical explanation 2]." We can directly obtain information from the original label for normal/abnormal and specific diseases or conditions. In addition, we have designed a series of rules based on specific diseases or conditions to help generate corresponding clinical explanations, as shown below:

1. Abnormal, Mild Non-Proliferative Diabetic Retinopathy, Only microaneurysms observed.
2. Abnormal, Moderate Non-Proliferative Diabetic Retinopathy, Retinal hemorrhages or hard exudates observed.
3. Abnormal, Severe Non-Proliferative Diabetic Retinopathy, Many intraretinal hemorrhages or definite venous beading observed.
4. Abnormal, Proliferative Diabetic Retinopathy, Neovascularization or vitreous/preretinal hemorrhage.
5. Abnormal, Glaucoma, Abnormal optic disk color and unclear optic disk boundaries.
6. Abnormal, Cataract, Opacification of crystalline lens observed, Abnormal fundus color.
7. Abnormal, Hypertensive Retinopathy, Abnormal arterial vein ratio, Abnormal fundus color.
8. Abnormal, Myopia, Leopard fundus observed, Abnormal fundus color.
9. Abnormal, Media Haze, Opacity of media observed, Abnormal fundus color.
10. Abnormal, Drusen, Yellow or white extracellular deposits located between the retinal pigment epithelium (RPE) and Bruch’s membrane, Abnormal fundus color.
11. Abnormal, Branch Retinal Vein Occlusion, Occlusion of the central retinal vein, Abnormal fundus color.
12. Abnormal, Tessellation, The choroidal vessels are visible due to the reduced density of the pigments, Abnormal fundus color.
13. Abnormal, Epiretinal Membrane, A thin fibrous or cellular membrane that forms on the inner surface of the retina, Abnormal fundus color.
14. Abnormal, Laser Scars, Circular or irregular shaped scars on the retinal surface observed, Abnormal fundus color.
15. Abnormal, Macular Scar, Scar on the macula observed, Abnormal fundus color.
16. Abnormal, Central Serous Retinopathy, Fluid accumulation under the retina observed, Abnormal fundus color.
17. Abnormal, Optic Disk Cupping, The thinning of neuroretinal rim such that optic disc appears excavated, Abnormal fundus color.
18. Abnormal, Central Retinal Vein Occlusion, Occlusion of the central retinal vein, The presence of flame-shaped hemorrhages, Abnormal fundus color.
19. Abnormal, Tortuous Vessels, Marked tortuosity of the retinal blood vessels, Abnormal fundus color.
20. Abnormal, Asteroid Hyalosis, Numerous astroid bodies are dispersed in vitreous, Abnormal fundus color.
21. Abnormal, Optic Disc Pallor, Pale yellow discoloration of the optic disc, as well as absence of many small vessels, Abnormal fundus color.
22. Abnormal, Optic Disc Edema, Swelling of the optic disc.
23. Abnormal, Optociliary Shunt, Presence of prepapillary vascular loops or optociliary shunt vessels.
24. Abnormal, Anterior Ischemic Optic Neuropathy, Optic disc swelling and pallor.
25. Abnormal, Parafoveal Telangiectasia, Yellow, lipid-rich exudation or parafoveal graying or tortuous blood vessels.
26. Abnormal, Retinal Traction, Presence of traction and retinal traction detachment.
27. Abnormal, Retinitis, Presence of vitreous inflammation or intraretinal hemorrhage.
28. Abnormal, Chorioretinitis, The hard exudates observed.
29. Abnormal, Exudation, Retinal detachment.
30. Abnormal, Retinal Pigment Epithelium Changes, The structural changes in the RPE.
31. Abnormal, Macular Hole, A small retinal break located in the center of the fovea observed.
32. Abnormal, Retinitis Pigmentosa, The presence of bone-spicule deposits and arterial narrowing.
33. Abnormal, Cotton Wool Spots, The presence of soft exudates.
34. Abnormal, Coloboma, The missing of portion of tissue in both the choroid and retina.
35. Abnormal, Optic Disc Pit Maculopathy, The presence of optic disc pit.
36. Abnormal, Preretinal Hemorrhage, Boat-shaped hemorrhage which obscures the underlying retina.
37. Abnormal, Myelinated Nerve Fibers, Gray-white opaque lesions with feathery edges observed.
38. Abnormal, Hemorrhagic Retinopathy, The presence of flame-shaped hemorrhages.
39. Abnormal, Central Retinal Artery Occlusion, The presence of pale, whitening, and retinal swelling.
40. Abnormal, Tilted Disk, The tilting presence of the oval optic disc.
41. Abnormal, Cystoid Macular Edema, The presence of multiple cystoid areas in the macula and causes retinal edema. 
42. Abnormal, Post-traumatic Choroidal Rupture, The breaks in the choroid, Bruch’s membrane, and RPE. 
43. Abnormal, Choroidal Folds, The presence of folds in the choroid.
44. Abnormal, Vitreous Hemorrhage, The presence of extravasated blood in one of the spaces created around the vitreous body.
45. Abnormal, Macroaneurysm, Fusiform or round dilation of the retinal arterioles which occur in the temporal retina observed.
46. Abnormal, Vasculitis, The presence of inflammation of retinal blood vessels.
47. Abnormal, Branch Retinal Artery Occlusion, The presence of acute retinal artery obstructions.
48. Abnormal, Plaque, The plaque is present in retina.
49. Abnormal, Hemorrhagic Pigment Epithelial Detachment, The presence of hemorrhage from the Bruch’s membrane.
50. Abnormal, Collateral, New retinal vessels developed within the framework of existing vessel network.
51. Abnormal, Choroidal Neovascularization, The presence of subretinal fluid.
52. Abnormal, Cysticercosis, The presence of retinal edema and hemorrhage.
53. Abnormal, Giant Retinal Tear, The presence of retinal detachment and circumferential full-thickness tears of the retina.
54. Abnormal, Macular Edema, The macula region exhibits radially oriented cystoid pockets.
55. Abnormal, Optic Neuritis, The presence of optic disc swelling.
56. Abnormal, Retinal Detachment, The retina detaches from the retinal pigment epithelium.
57. Abnormal, Retinal Holes, small tears in the retina observed.
58. Abnormal, Retinal Tears, small breaches in the retina observed.
59. Abnormal, Hypertensive Retinopathy, Retinal hemorrhages or hard exudates observed.
60. Abnormal, Idiopathic Intracranial Hypertension, The presence of papilledema.
61. Normal, Healthy, Normal optic disk color and clear optic disk boundaries, Normal Macular color, Normal fundus color, No apparent retinopathy.

\textbf{Dialogue Automatic Generation:}

The InternLM~\cite{team2023internlm} and GPT-4~\cite{achiam2023gpt} are used to generate corresponding three rounds of dialogue data based on text descriptions. The prompt we use is as follows: "You will be provided with information about fundus images, including whether they are abnormal, optional specific disease, and diagnostic criteria. The fundus information is as follows: [Keyword]. You are an ophthalmologist who hopes to provide high-quality medical answers for patients. You need to provide the questions as the patients asking for fundus. The questions should be asking for more details and reasons. And you also need to provide a more detailed answer as an ophthalmologist. The answer should be appropriate to the question and the answer should be less than 200 words. The number of questions should be three." 
The text description of each image is filled as the keyword of the above prompt. After data cleaning, we can obtain three rounds of dialogue data corresponding to each fundus image, which can be used as the dialogue part of the MMFundus dataset.

\begin{figure*}[!t]
\setlength{\abovecaptionskip}{0mm}
\centerline{\includegraphics[width=\linewidth]{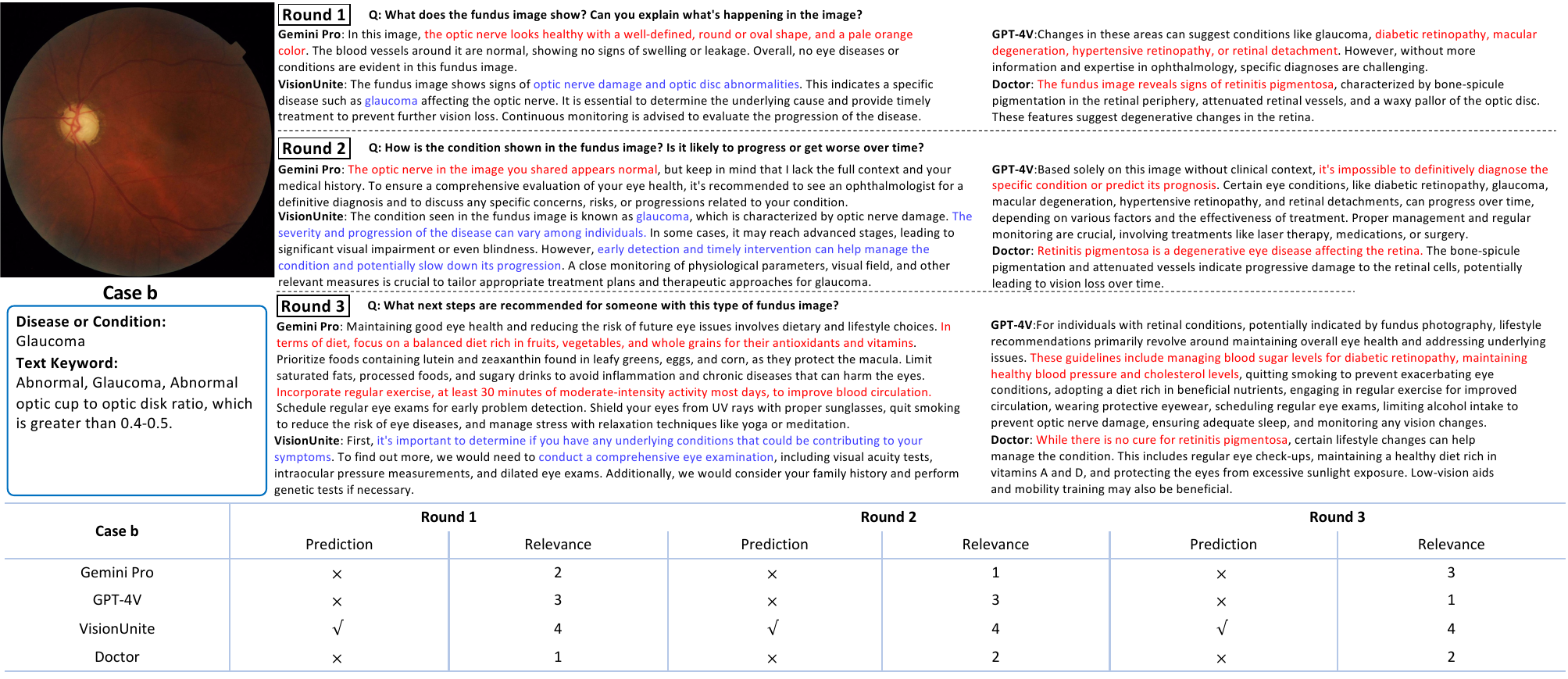}}
\caption{The consistent interpretation analysis of visual and language features between large vision-language models and the junior ophthalmologist (Doctor in the figure). \textcolor{red}{Red} represents inaccurate or irrelevant descriptions. \textcolor{blue}{Blue} represents the accurate descriptions. The expert (senior ophthalmologist) evaluation is also provided.}
\label{Fig6}
\vspace{-3mm}
\end{figure*}

\subsection{Implementation Details}
VisionUnite is implemented using Python 3.8.18 and Pytorch '2.0.1+cu117'. The fine-tuning of VisionUnite is run on 8-card NVIDIA A100 80G and 128 Intel(R) Xeon(R) Platinum 8369B CPU @ 2.90GHz. The batchsize we set is 4 per GPU in each mini-barch and the accumulated gradient iterations is 1, so the VisionUnite is trained with a batch size of 32. We utilize the AdamW optimizer with base lr=0.001 and the betas in AdamW is set as (0.9, 0.95). The absolute lr, which equals $baselr*batchsize/256=1.25E-4$. The lower lr bound for cyclic schedulers is 0 and the weight decay is 0.02. The number of training epochs is 10 in the pre-training stage. The number of training epochs is 30 in the fine-tuning stage and the first epochs are set for warming up. The number of workers we set is 10. The max token length is 512 to ensure that all the text can be included. We adopt the DistributedDataParallel (DDP) as the data parallel mechanism during training.

\subsection{Data Preprocessing}
In the curation and refinement of pre-training datasets, we initially gather over 1.6 million image-text pairs from the PMC-OA~\cite{PMCOA2003, lin2023pmc} and Retina Image Bank datasets~\cite{RetinaBank}. However, the PMC-OA dataset contained numerous low-quality non-biomedical images, including instances such as table images. Therefore, a meticulous screening and processing approach is undertaken. A ResNet18-based modality classifier is trained to discern non-biomedical images, with modalities defined as CT (Computed Tomography), FA (Fluorescein Angiography), Fundus, MRI (Magnetic Resonance Imaging), OCT (Optical Coherence Tomography), Pathology, PET (Positron Emission Tomography), X-ray, and Table Chart. Leveraging the predicted classification results, we retain images identified as CT, FA, Fundus, MRI, OCT, Pathology, PET, and X-ray, while discarding those with lower classification confidence. Concurrently, we refine the text component by removing words such as arrow, line, star, and various colors (red, yellow, blue, orange, green, purple, violet, black, white, and gray) to minimize interference from irrelevant information. Furthermore, we introduce explicit modality indicators in the image captions, such as "This is a Fundus image." This kind of deliberate inclusion of clear modality indicators aims to equip VisionUnite with the capability to analyze modalities during the pre-training stage.
To augment the fine-tuning dataset (MMFundus), we employ the ImageEnhance Contrast method to adjust the contrast of images. It involves creating an enhancement object and utilizing the enhancement method to yield an enhanced image. The enhancement factor (a floating-point parameter) governs the degree of enhancement: an enhancement factor of 0.0 generates a pure gray image, 1.0 preserves the original image, and higher values intensify image contrast. In our model, we set the enhancement factor to 1.3. Additionally, we perform a color space adjustment, transitioning from RGB to HSV. The HSV space is better suited for image processing as it segregates color and brightness information, thereby mitigating the impact of shadows and other extraneous factors.

\begin{figure*}[!t]
\setlength{\abovecaptionskip}{0mm}
\centerline{\includegraphics[width=\linewidth,height=24.5cm]{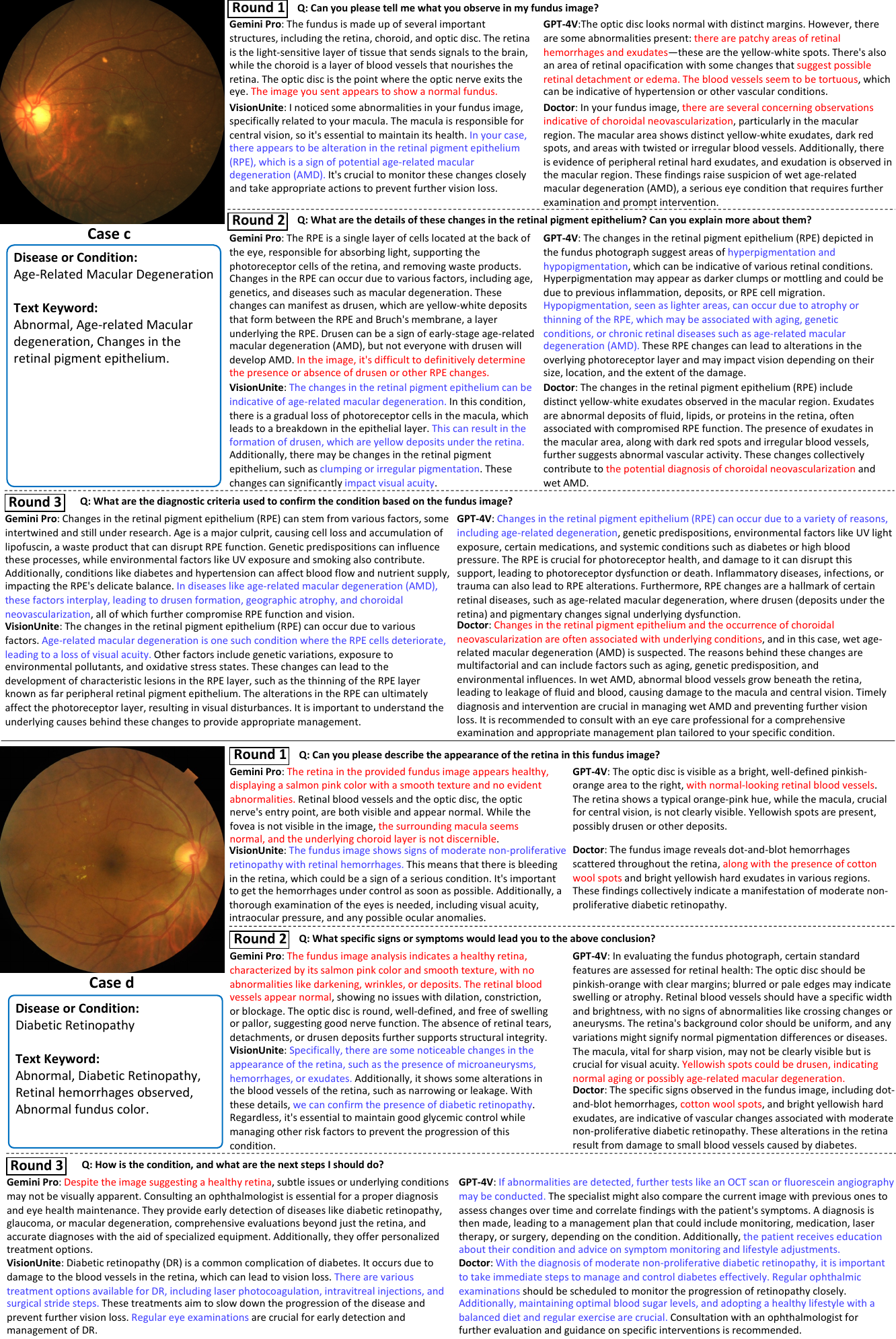}}
\caption{The consistent interpretation analysis of case c and case d with the single Disease or Condition (age-related macular degeneration and diabetic retinopathy) between large vision-language models and the junior ophthalmologist (Doctor in the figure). \textcolor{red}{Red} represents inaccurate or irrelevant descriptions. \textcolor{blue}{Blue} represents the accurate descriptions.}
\label{ExtendedFig6}
\end{figure*}

\begin{figure*}[!t]
\setlength{\abovecaptionskip}{0mm}
\centerline{\includegraphics[width=\linewidth]{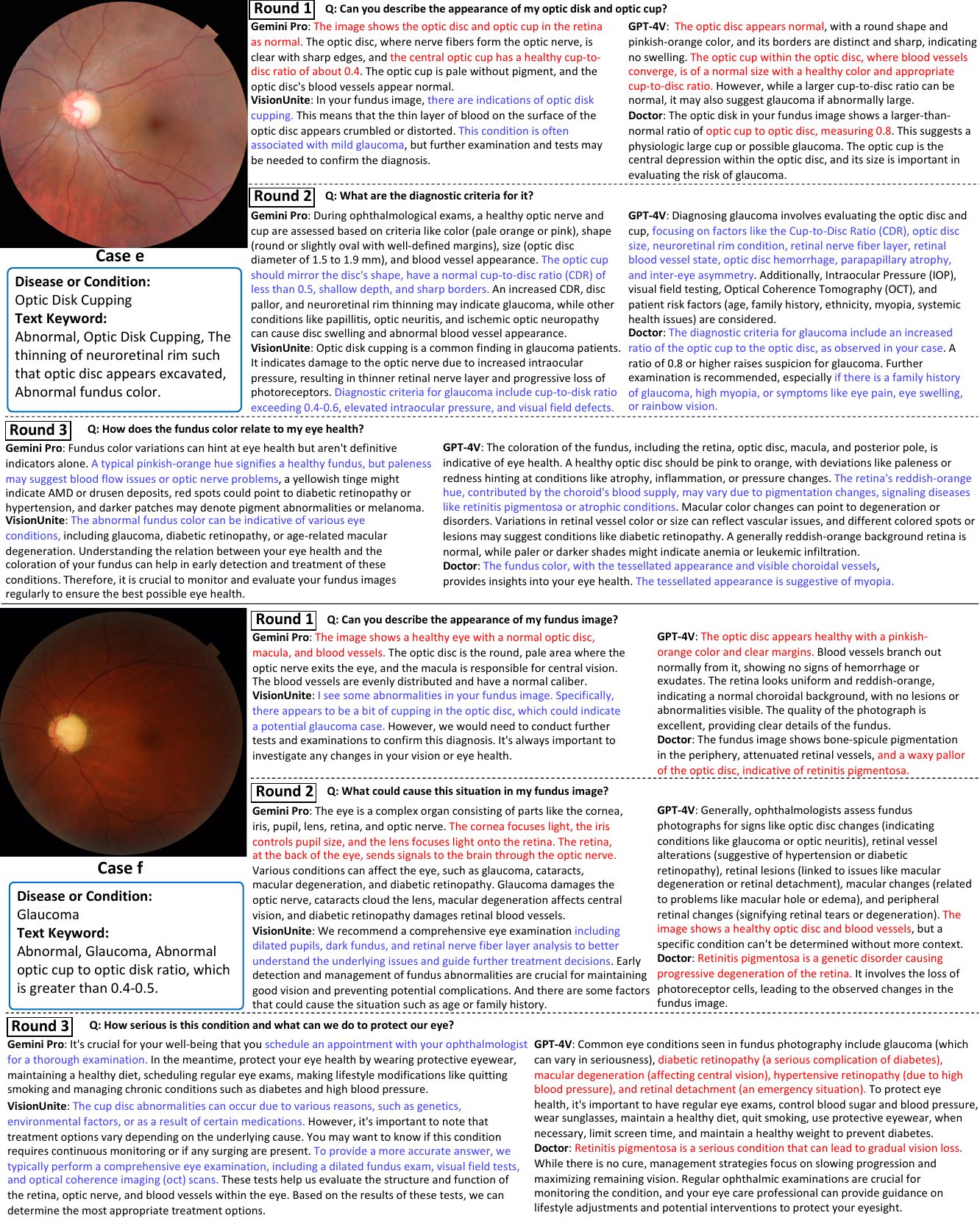}}
\caption{\textcolor{Revision}{The consistent interpretation analysis of case e and case f with the single Disease or Condition (optic disk cupping and glaucoma) between large vision-language models and the junior ophthalmologist (Doctor in the figure).} \textcolor{red}{Red} represents inaccurate or irrelevant descriptions. \textcolor{blue}{Blue} represents the accurate descriptions.}
\label{ExtendedFig7}
\end{figure*}

\begin{figure*}[!t]
\setlength{\abovecaptionskip}{0mm}
\centerline{\includegraphics[width=\linewidth,height=24.5cm]{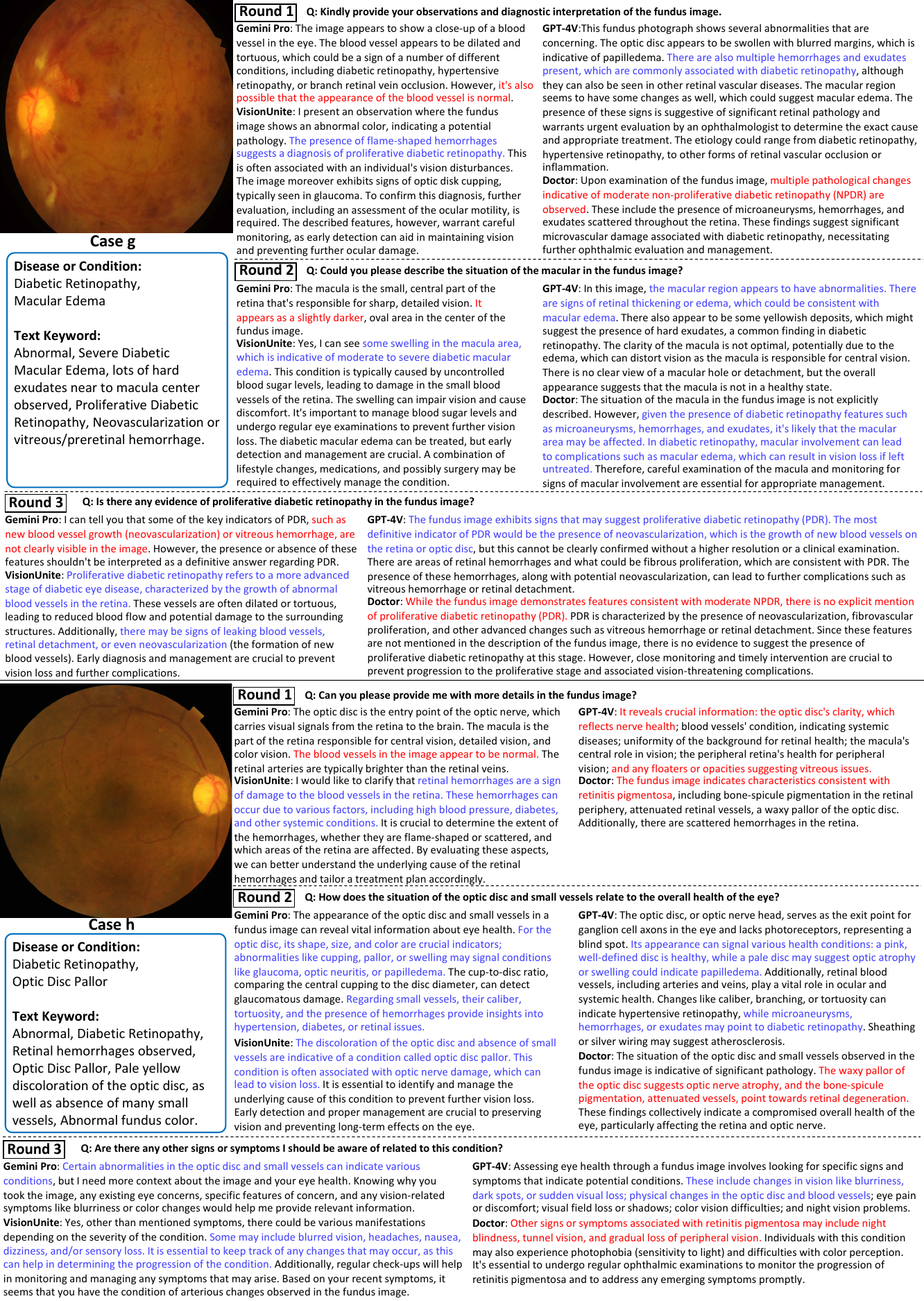}}
\caption{The consistent interpretation analysis of case g and case h with multiple Diseases or Conditions (diabetic retinopathy/macular edema and diabetic retinopathy/optic disc pallor) between large vision-language models and the junior ophthalmologist (Doctor in the figure). \textcolor{red}{Red} represents inaccurate or irrelevant descriptions. \textcolor{blue}{Blue} represents the accurate descriptions.}
\label{ExtendedFig8}
\end{figure*}

\begin{figure*}[!t]
\setlength{\abovecaptionskip}{0mm}
\centerline{\includegraphics[width=\linewidth,height=24.5cm]{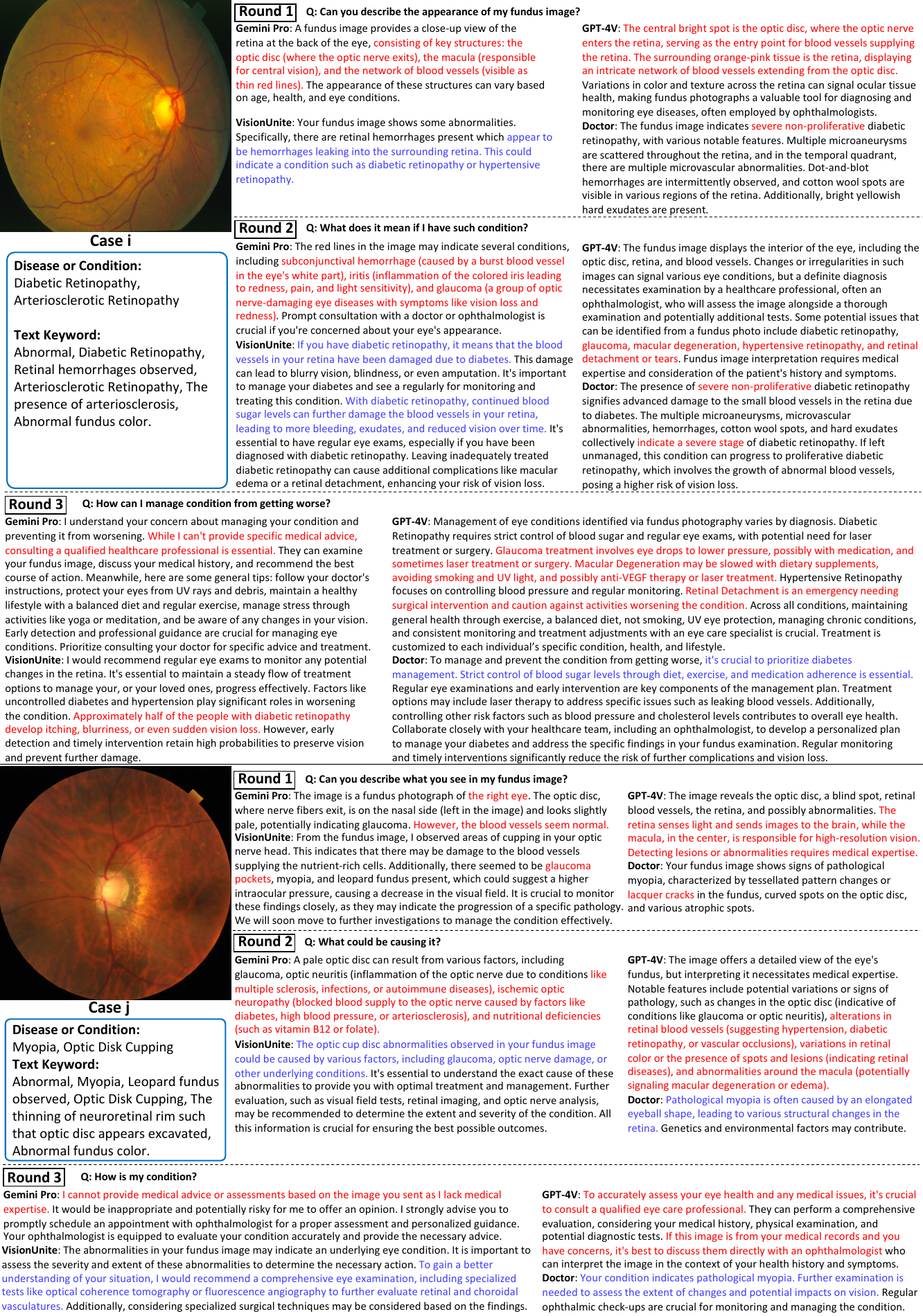}}
\caption{The consistent interpretation analysis of case i and case j with multiple Diseases or Conditions (diabetic retinopathy/arteriosclerotic retinopathy and myopia/optic disk cupping) between large vision-language models and the junior ophthalmologist (Doctor in the figure). \textcolor{red}{Red} represents inaccurate or irrelevant descriptions. \textcolor{blue}{Blue} represents the accurate descriptions.}
\label{ExtendedFig9}
\end{figure*}

\begin{figure*}[!t]
\setlength{\abovecaptionskip}{0mm}
\centerline{\includegraphics[width=\linewidth]{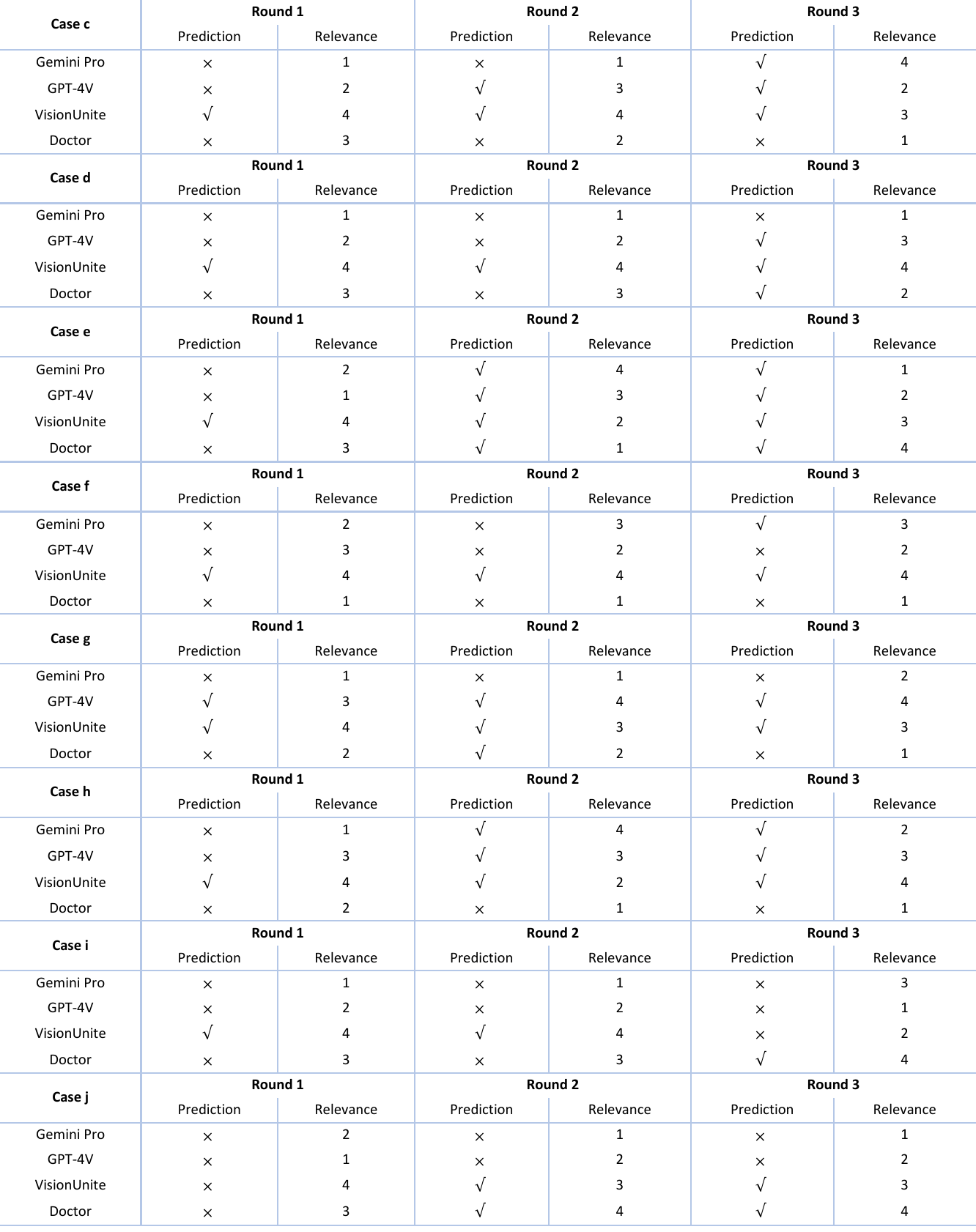}}
\caption{The corresponding expert (senior ophthalmologists) evaluation of case c-j, which includes the evaluation of diagnostic predictions and their relevance. $\times$ means incorrect prediction. $\checkmark$ means correct prediction. The expert assesses the relevance of each response set for diagnosis by ranking them from 4 to 1, based on their alignment with the label. A ranking of 4 indicates the highest consistency with the label, while a ranking of 1 indicates the lowest.}
\label{ExtendedFig10}
\end{figure*}

\bibliographystyle{IEEEtran}
\bibliography{main}